\documentclass[journal,twoside]{IEEEtran}
\pdfoutput=1
\usepackage{graphicx}
\usepackage{newtxtext}
\usepackage{newtxmath}
\usepackage{amsmath}
\usepackage{bm}
\usepackage{cite}
\usepackage[hidelinks]{hyperref}
\usepackage{eso-pic}
\usepackage[font=footnotesize]{subfig}
\usepackage{xcolor}
\usepackage{booktabs}
\usepackage{flushend}

\graphicspath{{./figures/}}

\newcommand{\Vector}[1]{\bm{#1}}  
\newcommand{\Matrix}[1]{\bm{#1}}  
\newcommand{\Transpose}{\mathrm{T}}  
\newcommand{\refEq}[1]{(\ref{#1})}               
\newcommand{\refEqBegin}[1]{Equation (\ref{#1})} 
\newcommand{\refFig}[1]{Fig.~\ref{#1}}           
\newcommand{\refSec}[1]{Sect.~\ref{#1}}          
\newcommand{\refTable}[1]{Table~\ref{#1}}        
\newcommand{\refTableBegin}[1]{Table~\ref{#1}}   

\hypersetup{%
    pdfauthor={Gernot Herbst and Rafal Madonski},
    pdftitle={Tuning and Implementation Variants of Discrete-Time ADRC},
    pdfcreator={},pdfproducer={}
}

\begin{document}

\AddToShipoutPicture*{%
  \AtPageUpperLeft{%
    \setlength\unitlength{1cm}%
    \put(0,-0.5){\begin{minipage}[c]{\paperwidth}
    \footnotesize\centering\textcolor{black!50}{%
    This is a preprint of an article published in \emph{Control Theory and Technology}.
    The final authenticated version is available online at:} \ \textcolor{blue!60}{\url{https://doi.org/10.1007/s11768-023-00127-0}}%
    \end{minipage}}%
  }
}

\title{Tuning and Implementation Variants of Discrete-Time ADRC}

\author{Gernot Herbst\ \href{https://orcid.org/0000-0002-4638-5378}%
    {\raisebox{-0.3pt}{\includegraphics[height=9pt]{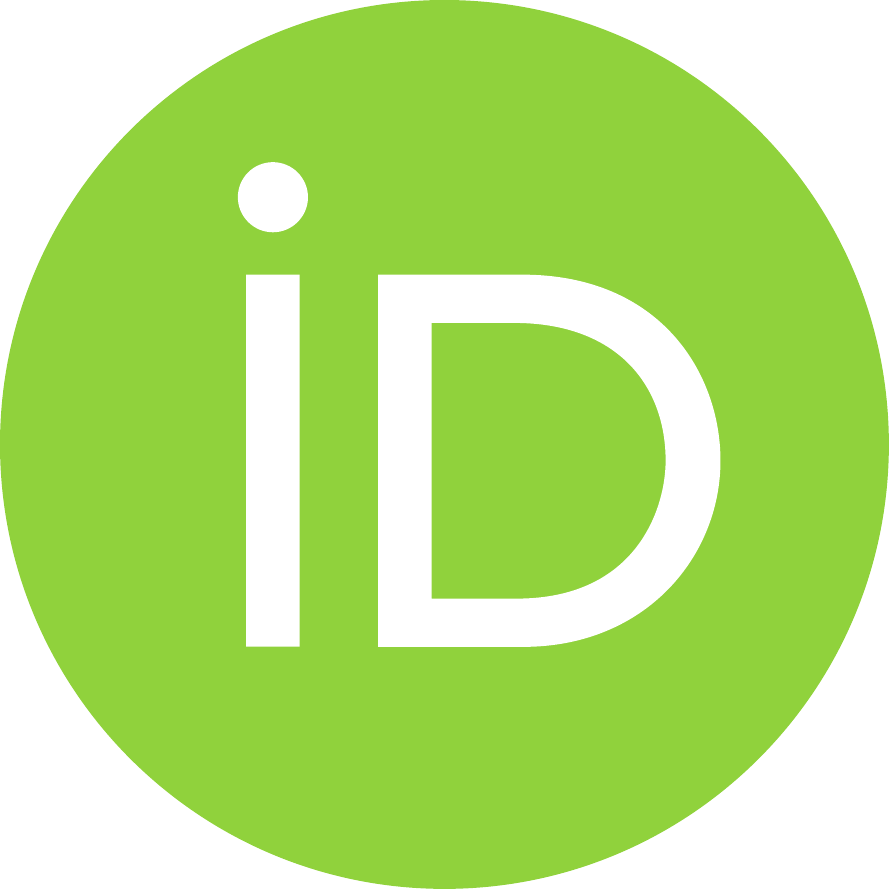}}}%
  \thanks{%
    Gernot Herbst is with University of Applied Sciences Zwickau, Germany (e-mail: \protect\href{mailto:gernot.herbst@fh-zwickau.de}{gernot.herbst@fh-zwickau.de}).%
  }
  and
  Rafal Madonski\ \href{https://orcid.org/0000-0002-1798-0717}%
    {\raisebox{-0.3pt}{\includegraphics[height=9pt]{orcid}}}%
  \thanks{%
    Rafal Madonski is with the Energy and Electricity Research Center, Jinan University, China (e-mail: \protect\href{mailto:rafal.madonski@jnu.edu.cn}{rafal.madonski@jnu.edu.cn}).%
  }
}

\markboth%
{Gernot Herbst and Rafal Madonski: Tuning and Implementation Variants of Discrete-Time ADRC}
{Gernot Herbst and Rafal Madonski: Tuning and Implementation Variants of Discrete-Time ADRC}

\maketitle

\begin{abstract}
    Practical implementations of active disturbance rejection control (ADRC) will almost always take place in discretized form.
    Since applications may have quite different needs regarding their discrete-time controllers, this article summarizes and extends the available set of ADRC implementations to provide a suitable variant for as many as possible use cases.
    In doing so, the gap between quasi-continuous and discrete-time controller tuning is being closed for applications with low sampling frequencies.
    The main contribution of this article is the derivation of three different discrete-time implementations of error-based ADRC.
    It is shown that these are almost one-to-one counterparts of existing output-based implementations, to the point where transfer functions and coefficients can be reused in unaltered form.
    In this way, error-based implementations become firmly rooted in the established landscape of discrete-time ADRC.
    Furthermore, it becomes possible to equip error-based variants with windup protection abilities known from output-based ADRC.
\end{abstract}



\section{Introduction}

\subsection{Background and Related Work}

Active disturbance rejection control (ADRC) \cite{Han:2009} is a highly practical approach to solving real-world control problems.
As the dominance of PID controller families in this domain is undisputed despite decades of control systems research, the widespread interest ADRC has sparked among both scientists and engineers is all the more remarkable, since it already found its way into numerous applications \cite{Zheng:2010,Zheng:2018,Talole:2018,Farah:2021}.
Some foundational milestones enabling this trajectory deep into the territory of PID control can be seen in the development of a linear version \cite{Gao:2006}, an easy tuning rule \cite{Gao:2003}, and its implementation in discrete-time form \cite{Miklosovic:2006}.
Later enhancements include additional features for industrial practice \cite{Madonski:2015,Herbst:2016a}, methods for improved disturbance rejection \cite{Madonski:2015b}, or the incorporation of known plant information \cite{Fu:2016,Zhou:2019}.

An important barrier to be overcome for real-world adoption is not a technical one, but more a matter of understanding.
With its origins in (nonlinear) state space, ADRC might appear unfamiliar to the practicing engineer.
To cater such needs, a multitude of studies provides insights and understanding of ADRC from a classical, frequency-domain control engineering perspective \cite{Tian:2007,Huang:2013:CCC,Zheng:2016,Herbst:2021a}.

Taking this avenue also for the actual implementation---i.\,e.\ building ADRC using transfer functions---not only eases understanding, but increases efficiency and allows to even reuse existing building blocks from classical control engineering.
For discrete-time ADRC, this was put forward by \cite{Herbst:2021a}, with an even more optimized variant in \cite{Herbst:2021b} that comes very close to the runtime performance of PID controllers.
These are implementations we will build upon in this article.

A variant of ADRC, using the control error as estimated variable within the observer, was introduced in  \cite{Michalek:2016,Mandali:2020}, and accordingly named \emph{error-based ADRC}.
When viewed in transfer function form, it becomes apparent that error-based ADRC is a one-degree of freedom (1DOF) controller, as only one transfer function is needed---a feedback controller in a structure very well known from classical control engineering.
Applications studied so far include quadrocopters \cite{Lechekhab:2021}, DC-DC converters \cite{Madonski:2021}, and thermal processes \cite{Huang:2022}.
The firm connection between output- and error-based ADRC has been established in continuous-time domain \cite{Madonski:2023a}.
To the best of our knowledge, however, the task of discretization of error-based ADRC has not yet been properly addressed, apart from an approximate solution using bilinear transform in \cite{Madonski:2019}.
We therefore want to extend this connection to the discrete-time domain in this article, providing ready-to-use variants that are exactly rooted in the discretization process of ADRC introduced by \cite{Miklosovic:2006}.

\subsection{Contributions and Structure of this Article}

This article aims at increasing the versatility of discrete-time ADRC in two aspects.
Firstly, tuning rules for the outer state-feedback controller based on bandwidth parameterization \cite{Gao:2003} with discrete-time pole placement are introduced, which are needed to consistently achieve the desired bandwidth, even for applications with low sampling frequencies.
At the same time, a bridge is being built to the commonly used continuous-time design.
Existing transfer function implementations of discrete-time ADRC from \cite{Herbst:2021a} and \cite{Herbst:2021b} are being reviewed and retrofitted with these results.
Secondly, discrete-time state-space and transfer function implementations of error-based ADRC \cite{Madonski:2019} are developed.
These exhibit strong connections with classical (output-based) variants up to equivalent transfer function coefficients, and enhance error-based ADRC with flexible control signal limitation and windup protection.

The remainder of this article is structured as follows:
\refSec{sec:ADRC_CT} briefly recapitulates continuous-time ADRC, which we will need for deriving the discrete-time implementations.
Building on that, \refSec{sec:ADRC_DT} summarizes the current practice of implementing discrete-time ADRC in state space form.
As will be shown, it relies on quasi-continuous tuning of the outer feedback controller within ADRC.
\refSec{sec:DT_Tuning} will therefore put the commonly used bandwidth parameterization approach of the controller on the basis of discrete-time pole placement.
Two existing transfer function implementations of discrete-time ADRC are revisited in \refSec{sec:DT_TF}, now providing tuning equations based on the results of \refSec{sec:DT_Tuning}.
For error-based ADRC, three different implementations are introduced in \refSec{sec:ErrorBased}, all as direct counterparts of existing output-based variants.
The article concludes after presenting an example in \refSec{sec:Example} that compares output- and error-based ADRC in a power electronics control application.

\section{Summary of Continuous-Time ADRC}
\label{sec:ADRC_CT}

\subsection{Structure and Control Law}
\label{sec:ADRC_CT_ControlLaw}

\begin{figure}
    \includegraphics[width=\linewidth]{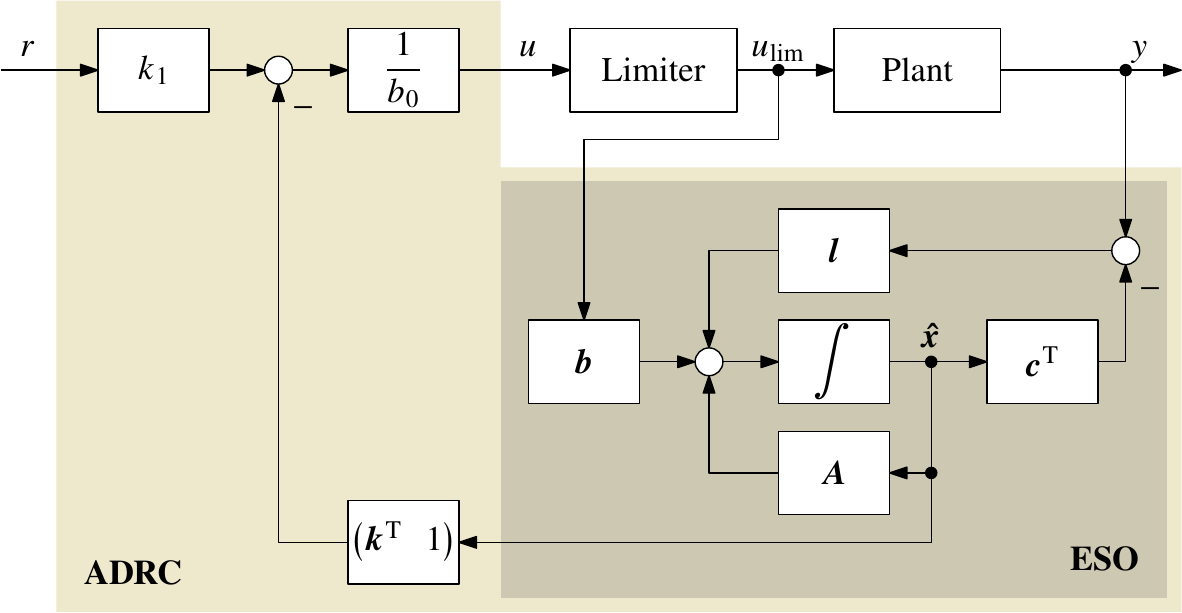}
    \caption{Continuous-time ADRC in state space, with an external, user-defined control signal limiter}
    \label{fig:ADRC_CT_SS}
\end{figure}

All discrete-time implementations discussed in this article are based on the linear variant of ADRC introduced in \cite{Gao:2003}, which we consider the most widely used form.
We will briefly summarize it in this section.
The control law of continuous-time linear ADRC in state space as shown in \refFig{fig:ADRC_CT_SS} reads:
\begin{gather}
    u(t) = \frac{1}{b_0} \cdot \left( k_1 \cdot r(t) - \begin{pmatrix}
        \Vector{k}^\Transpose  &  1
    \end{pmatrix} \cdot \Vector{\hat{x}}(t) \right)
    \label{eqn:ADRC_CT_SS_Controller}
    \\
    \text{with}\quad
    \Vector{k}^\Transpose =
    \begin{pmatrix}
        k_1  &  \cdots  &  k_n
    \end{pmatrix}
    \quad\text{and}\quad
    \Vector{\hat{x}} = \begin{pmatrix}
        \hat{x}_1  &  \cdots  &  \hat{x}_{n+1}
    \end{pmatrix}^\Transpose
    \notag
    .
\end{gather}

In \refEq{eqn:ADRC_CT_SS_Controller}, the unity-gain factor in the state-feedback gain vector corresponds to the rejection of the total disturbance.
The vector of states $\Vector{\hat{x}}(t)$ is being provided by a Luenberger observer.
Its plant model most commonly consists of an integrator chain disturbed by a constant input disturbance (the total disturbance $f(t)$): $y^{(n)}(t) = b_0 \cdot u(t) + f(t)$.
That way, only the plant order $n$ and an estimated plant gain $b_0$ are required as far as modeling is concerned.
The observer equations are therefore given as:
\begin{gather}
    \Vector{\dot{\hat{x}}}(t)
    = \Matrix{A} \cdot \Vector{\hat{x}}(t)
    + \Vector{b} \cdot u_\mathrm{lim}(t)
    + \Vector{l} \cdot \left( y(t) - \Vector{c}^\Transpose \cdot  \Vector{\hat{x}}(t) \right)
    \label{eqn:ADRC_CT_SS_Observer}
    \\
    \text{with}\quad
    \Matrix{A} = \begin{pmatrix}
        \Matrix{0}^{n \times 1}  &  \Matrix{I}^{n \times n}  \\
        0  &  \Matrix{0}^{1 \times n}
    \end{pmatrix}
    ,\quad
    \Vector{b} = \begin{pmatrix}
        \Matrix{0}^{(n-1) \times 1}  \\
        b_0  \\
        0
    \end{pmatrix}
    ,\notag\\
    \Vector{l} = \begin{pmatrix}
        l_1  &  \cdots  &  l_{n+1}
    \end{pmatrix}^\Transpose
    ,\quad
    \Vector{c}^\Transpose = \begin{pmatrix}
        1  &  \Matrix{0}^{1 \times n}
    \end{pmatrix}
    \notag
    .
\end{gather}

Note that in \refEq{eqn:ADRC_CT_SS_Observer}, the observer uses the limited control signal $u_\mathrm{lim}(t)$ instead of $u(t)$, which is a very simple but effective method of avoiding integrator windup \cite{Herbst:2016a,AstromMurray:2021}---a very important selling point for using ADRC in practice.
That way, no further anti-windup measures have to be implemented, let alone to be tuned.
Finally, it might also be possible that the control signal limitation takes place externally, beyond the user's influence.
For these cases, $u_\mathrm{lim}(t)$ provides a feedback path for a measured value of the actual (limited) control signal.

\subsection{Controller Tuning}
\label{sec:ADRC_CT_Controller_Tuning}

We will now briefly discuss tuning the gain vector $\Vector{k}^\Transpose$ of the outer state-feedback controller from \refEq{eqn:ADRC_CT_SS_Controller}.
This will also be the basis for developing discrete-time tuning rules in \refSec{sec:DT_Tuning_Bandwidth}.
The $1 / b_0$ gain block in \refFig{fig:ADRC_CT_SS} normalizes the plant gain, which, together with feeding back the estimated total disturbance, approximately creates an undisturbed ``virtual plant'' with unity-gain integrator chain dynamics $y^{(n)}(t) = u_0(t)$ to be controlled by the controller $u_0(t) = -\Vector{k}^\Transpose \cdot \begin{pmatrix} \hat{x}_1 & \cdots & \hat{x}_{n} \end{pmatrix}^\Transpose$.
A state space representation of this virtual plant can be given as:
\begin{gather}
    \begin{aligned}
        \Vector{\dot{x}}_\mathrm{VP}(t)
        &= \Matrix{A}_\mathrm{VP} \cdot \Vector{x}_\mathrm{VP}(t)
        + \Vector{b}_\mathrm{VP} \cdot u_0(t)
        ,\\
        y(t) &= \Vector{c}_\mathrm{VP}^\Transpose \cdot \Vector{x}_\mathrm{VP}(t)
    \end{aligned}
    \label{eqn:ADRC_CT_VirtualPlant}
    \\
    \text{with}\quad
    \Matrix{A}_\mathrm{VP} = \begin{pmatrix}
        \Matrix{0}^{(n-1) \times 1}  &  \Matrix{I}^{(n-1) \times (n-1)}  \\
        0  &  \Matrix{0}^{1 \times (n-1)}
    \end{pmatrix}
    ,\notag\\
    \Vector{b}_\mathrm{VP} = \begin{pmatrix}
        \Matrix{0}^{(n-1) \times 1}  \\
        1
    \end{pmatrix}
    ,\quad
    \Vector{c}_\mathrm{VP}^\Transpose = \begin{pmatrix}  1  &  \Matrix{0}^{1 \times (n-1)}  \end{pmatrix}
    \notag
    .
\end{gather}

The most popular tuning method for linear ADRC was introduced in \cite{Gao:2003} as \emph{bandwidth parameterization}.
Its simplicity comes from a single tuning parameter, namely, the desired bandwidth $\omega_\mathrm{CL}$.
In this approach, all poles of the closed loop are placed at a common location $-\omega_\mathrm{CL}$:
\begin{align}
    \left( \lambda + \omega_\mathrm{CL} \right)^n
    & \stackrel{!}{=}
    \det\left( \lambda \Matrix{I} - \left( \Matrix{A}_\mathrm{VP} - \Vector{b}_\mathrm{VP} \Vector{k}^\Transpose \right) \right)
    \label{eqn:ADRC_CT_SS_Controller_K_PolePlacement}
    \\
    & = \lambda^{n} + k_n \lambda^{n-1} + \ldots + k_2 \lambda + k_1
    .
    \notag
\end{align}

After binomal expansion, the controller gains $k_i$ can be read off \refEq{eqn:ADRC_CT_SS_Controller_K_PolePlacement}:
\begin{equation}
    k_i = \frac{n!}{(n-i+1)! \cdot (i-1)!} \cdot \omega_\mathrm{CL}^{n-i+1}
    \quad \forall i = 1, \ldots, n
    .
    \label{eqn:ADRC_CT_SS_Controller_K}
\end{equation}

\refEqBegin{eqn:ADRC_CT_SS_Controller_K} yields the very simple controller tuning equations
$k_1 = \omega_\mathrm{CL}$ (for first-order ADRC), and $k_1 = \omega_\mathrm{CL}^2$, $k_2 = 2 \omega_\mathrm{CL}$ (for second-order ADRC).
We will come back to these in \refSec{sec:DT_Tuning_Approximation}.

\subsection{Observer Tuning}
\label{sec:ADRC_CT_Observer_Tuning}

In a similar fashion, tuning the observer can be done by placing the poles of the observer system matrix $\left( \Matrix{A} - \Vector{l} \Vector{c}^\Transpose \right)$.
Using bandwidth parameterization once more, an (obviously higher) bandwidth has to be chosen for the observer, which we will---following \cite{Herbst:2013}---express as $k_\mathrm{ESO} \cdot \omega_\mathrm{CL}$. The equation for pole placement then reads:
\begin{align}
    \left( \lambda + k_\mathrm{ESO} \omega_\mathrm{CL} \right)^{n+1}
    & \stackrel{!}{=}
    \det\left( \lambda \Matrix{I} - \left( \Matrix{A} - \Vector{l} \Vector{c}^\Transpose \right) \right)
    \label{eqn:ADRC_CT_SS_Observer_L_PolePlacement}
    \\
    & = \lambda^{n+1} + l_1 \lambda^{n} + \ldots + l_{n} \lambda + l_{n+1}
    .
    \notag
\end{align}

The observer gains can be obtained from \refEq{eqn:ADRC_CT_SS_Observer_L_PolePlacement} after binominal expansion as follows:
\begin{equation}
    l_i = \frac{(n+1)!}{(n-i+1)! \cdot i!} \cdot \left( k_\mathrm{ESO} \cdot \omega_\mathrm{CL} \right)^i
    \quad \forall i = 1, \ldots, n+1
    .
    \label{eqn:ADRC_CT_SS_Observer_L}
\end{equation}

\section{Discrete-Time ADRC}
\label{sec:ADRC_DT}

\subsection{Structure and Control Law}
\label{sec:ADRC_DT_ControlLaw}

As the control law \refEq{eqn:ADRC_CT_SS_Controller} of linear ADRC consists of static state feedback, its discrete-time counterpart is trivially obtained as:
\begin{gather}
    u(k) = \frac{1}{b_0} \cdot \left( k_1 \cdot r(k) - \begin{pmatrix}
        \Vector{k}^\Transpose  &  1
    \end{pmatrix}
    \cdot \Vector{\hat{x}}(k) \right)
    \label{eqn:ADRC_DT_SS_Controller}
    \\
    \text{with}\quad
    \Vector{k}^\Transpose = \begin{pmatrix}
        k_1  &  \cdots  &  k_n
    \end{pmatrix}
    .
    \notag
\end{gather}

\begin{figure}
    \includegraphics[width=\linewidth]{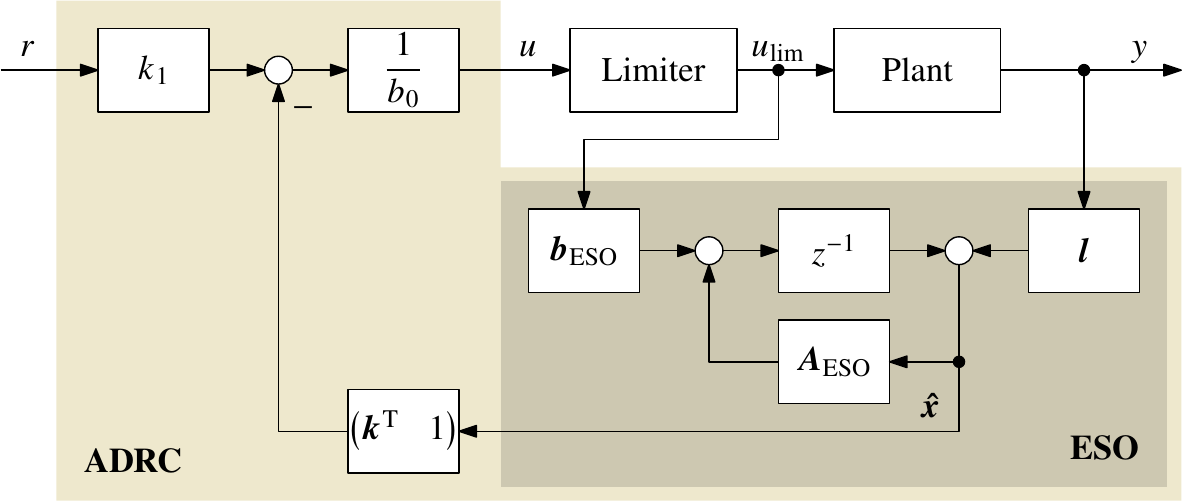}
    \caption{Control loop with discrete-time ADRC, state space implementation}
    \label{fig:ADRC_DT_SS_Limiter}
\end{figure}

When developing the discrete-time version of the extended state observer, the so-called \emph{current observer} approach is being used in order to let the current estimate $\Vector{\hat{x}}(k)$ be based on the most recent measurement $y(k)$ \cite{Franklin:1997,Miklosovic:2006}.
For a more detailed derivation we refer the reader to \cite{Herbst:2013}, and only present the resulting equations for the discrete-time observer dynamics here:
\begin{gather}
    \Vector{\hat{x}}(k) = \Matrix{A}_\mathrm{ESO} \cdot \Vector{\hat{x}}(k-1) + \Vector{b}_\mathrm{ESO} \cdot u_\mathrm{lim}(k-1) + \Vector{l} \cdot y(k)
    \label{eqn:ADRC_DT_SS_Observer}
    \\
    \text{with}\quad
    \Vector{l} = \begin{pmatrix}
        l_1  &  \cdots  &  l_{n+1}
    \end{pmatrix}^\Transpose
    ,\quad
    \Matrix{A}_\mathrm{ESO} = \Matrix{A}_\mathrm{d} - \Vector{l} \cdot \Vector{c}^\Transpose_\mathrm{d} \cdot \Matrix{A}_\mathrm{d}
    ,\notag\\
    \text{and}\quad
    \Vector{b}_\mathrm{ESO} = \Vector{b}_\mathrm{d} - \Vector{l} \cdot \Vector{c}^\Transpose_\mathrm{d} \cdot \Vector{b}_\mathrm{d}
    .
    \notag
\end{gather}

In \refEq{eqn:ADRC_DT_SS_Observer}, $\Matrix{A}_\mathrm{d}$, $\Vector{b}_\mathrm{d}$, and $ \Vector{c}^\Transpose_\mathrm{d}$ are obtained from zero-order hold (ZOH) discretization with a sampling interval $T$ of their respective counterparts in \refEq{eqn:ADRC_CT_SS_Observer}, as recommended by \cite{Miklosovic:2006}:
\begin{gather}
    \Matrix{A}_\mathrm{d} = \Matrix{I} + \sum_{i = 1}^\infty \displaystyle\frac{\Matrix{A}^i T^i}{i!}
    ,\ \
    \Vector{b}_\mathrm{d} = \left( \sum_{i = 1}^\infty \displaystyle\frac{\Matrix{A}^{i-1} T^i}{i!} \right) \cdot \Vector{b}
    ,\ \
    \Vector{c}^\Transpose_\mathrm{d} = \Vector{c}^\Transpose
    .
    \notag
\end{gather}

Controller \refEq{eqn:ADRC_DT_SS_Controller} and observer \refEq{eqn:ADRC_DT_SS_Observer} are shown as a block diagram in \refFig{fig:ADRC_DT_SS_Limiter}.
Once again, one can recognize ADRC's signature ingredients: extended state observer (built around a simple plant model with input disturbance), disturbance rejection (unity-gain feedback of the state $\hat{x}_{n+1}$), and plant gain inversion (multiplication by $1/b_0$ to obtain $u(k)$).

\subsection{Controller Tuning}
\label{sec:ADRC_DT_Controller_Tuning}

The outer state feedback controller of ADRC consists of a static feedback gain vector $\Vector{k}^\Transpose$, which means that there are no dynamics to be discretized for the controller.
Assuming that the sampling interval $T$ is negligible compared to the dynamics of the closed control loop, the continuous-time design from \refSec{sec:ADRC_CT_Controller_Tuning} can be employed with very good accuracy for a discrete-time implementation, as well.

And indeed, to the best of the authors' knowledge, this quasi-continuous controller design within a discrete-time linear ADRC implementation has, up to now, always been performed this way, see \cite{Miklosovic:2007,Zhang:2013,Stankovic:2016,Ahi:2018,Desai:2018,RamirezNeria:2020} for some examples.
For applications with relatively large sampling intervals, however, the quasi-continuous design does not exactly lead to the desired closed-loop behavior.
To round off ADRC controller tuning even for these edge cases, we therefore want to introduce discrete-time tuning rules for $\Vector{k}^\Transpose$ in this article, which will be presented in the \refSec{sec:DT_Tuning}.

\subsection{Observer Tuning}
\label{sec:ADRC_DT_Observer_Tuning}

Looking back at \refEq{eqn:ADRC_DT_SS_Observer}, it becomes obvious that the abbreviation ${A}_\mathrm{ESO}$ is the system matrix of the discrete-time observer.
Corresponding to the continuous-time case, we can use the pole placement approach of bandwidth parameterization for tuning the observer.
Before doing so, we need to map the desired common pole from its $s$-plane to the $z$-domain location $z_\mathrm{ESO}$:
\begin{gather}
    \det\left( z \Matrix{I} - \Matrix{A}_\mathrm{ESO} \right)
    \stackrel{!}{=} \left( z - z_\mathrm{ESO} \right)^{n+1}
    \label{eqn:ADRC_DT_SS_Observer_Approach}
    \\
    \text{with}\quad
    z_\mathrm{ESO} = \mathrm{e}^{- k_\mathrm{ESO} \cdot \omega_\mathrm{CL} \cdot T}
    .
    \notag
\end{gather}

As shown before, for example in \cite{Miklosovic:2006,Herbst:2013}, this results in the observer gains
$l_1 = 1 - z_\mathrm{ESO}^2$ and $l_2 = \frac{1}{T} \cdot \left( 1 - z_\mathrm{ESO} \right)^2$
for the first-order case, and three gain values
$l_1 = 1 - z_\mathrm{ESO}^3$, $l_2 = \frac{3}{2T} \cdot \left(1 - z_\mathrm{ESO}\right)^2 \cdot \left(1 + z_\mathrm{ESO}\right)$, $l_3 = \frac{1}{T^2} \cdot \left(1 - z_\mathrm{ESO}\right)^3$
for second-order ADRC.

\section{Discrete-Time Controller Tuning}
\label{sec:DT_Tuning}

\subsection{Bandwidth Parameterization}
\label{sec:DT_Tuning_Bandwidth}

In a discrete-time implementation of ADRC, all inputs and outputs are updated only once per sampling interval $T$.
The observer will, naturally, provide estimates for the outer state-feedback controller and the total disturbance only at these time instants, as well.
A discrete-time design of the controller $\Vector{k}^\Transpose$ therefore requires a discrete-time model of the virtual plant \refEq{eqn:ADRC_CT_VirtualPlant}, as well.
Assuming a zero-order hold behavior of the actuator, we can derive the following state-space model:
\begin{gather}
    \begin{aligned}
        \Vector{\hat{x}}_\mathrm{VP}(k+1) &= \Matrix{A}_\mathrm{VP,d} \cdot \Vector{\hat{x}}_\mathrm{VP}(k) + \Vector{b}_\mathrm{VP,d} \cdot u(k)
        ,\\
        y_\mathrm{VP}(k) &= \Vector{c}^\Transpose_\mathrm{VP,d} \cdot \Vector{\hat{x}}_\mathrm{VP}(k)
    \end{aligned}
    \label{eqn:ADRC_DT_VirtualPlant}
    \\
    \text{with}\quad
    \Matrix{A}_\mathrm{VP,d} = \Matrix{I} + \sum_{i = 1}^\infty \displaystyle\frac{\Matrix{A}_\mathrm{VP}^i T^i}{i!}
    ,\notag\\
    \Vector{b}_\mathrm{VP,d} = \left( \sum_{i = 1}^\infty \displaystyle\frac{\Matrix{A}_\mathrm{VP}^{i-1} T^i}{i!} \right) \cdot \Vector{b}_\mathrm{VP}
    ,\quad
    \Vector{c}^\Transpose_\mathrm{VP,d} = \Vector{c}_\mathrm{VP}^\Transpose
    .
    \notag
\end{gather}

Similar the observer tuning procedure described in \refSec{sec:ADRC_DT_Observer_Tuning}, we can now apply pole placement with bandwidth parameterization for the discrete-time controller.
Firstly, we map the desired location for all closed-loop poles from $s$- to $z$-domain:
\begin{equation}
    z_\mathrm{CL} = \mathrm{e}^{-\omega_\mathrm{CL} T}
    ,
    \label{eqn:ADRC_DT_zCL}
\end{equation}
and then the design equation will read:
\begin{equation}
    \det\left( z \Matrix{I} - \left( \Matrix{A}_\mathrm{VP,d} - \Vector{b}_\mathrm{VP,d} \Vector{k}^\Transpose \right) \right)
    \stackrel{!}{=} \left( z - z_\mathrm{CL} \right)^n
    .
    \label{eqn:ADRC_DT_SS_Controller_K_PolePlacement}
\end{equation}

The most relevant cases for a practical implementation are $n = 1$ and $n = 2$, where ADRC will be used to replace PI and PID controllers, respectively.
For the first-order case, $\Matrix{A}_\mathrm{VP,d}$ and $\Vector{b}_\mathrm{VP,d}$ are only scalars:
\begin{equation*}
    \Matrix{A}_\mathrm{VP,d} = \begin{pmatrix}  1  \end{pmatrix}
    \quad\text{and}\quad
    \Vector{b}_\mathrm{VP,d} = \begin{pmatrix}  T  \end{pmatrix}
    .
\end{equation*}

This means that \refEq{eqn:ADRC_DT_SS_Controller_K_PolePlacement} will be reduced to the following simple equation:
\begin{equation*}
    z + \left( T k_1 - 1 \right) \stackrel{!}{=} z - z_\mathrm{CL} \cdot
    ,
\end{equation*}
from which we can easily obtain the discrete-time controller tuning equation for $k_1$ as:
\begin{equation}
    k_1 = \frac{1 - z_\mathrm{CL} }{ T }
    .
    \label{eqn:ADRC_DT_SS_Controller_K_n1}
\end{equation}

In the second-order case, $\Matrix{A}_\mathrm{VP,d}$ and $\Vector{b}_\mathrm{VP,d}$ are:
\begin{equation*}
    \Matrix{A}_\mathrm{VP,d} = \begin{pmatrix}  1 \ & \ T \\ 0 \ & \ 1  \end{pmatrix}
    \quad\text{and}\quad
    \Vector{b}_\mathrm{VP,d} = \begin{pmatrix}  \frac{1}{2} T^2 \\ T  \end{pmatrix}
    .
\end{equation*}

The design equation \refEq{eqn:ADRC_DT_SS_Controller_K_PolePlacement} now becomes:
\begin{align*}
    &z^2 + \left( \frac{k_1 T^2}{2} + k_2 T - 2 \right) \cdot z + \left( \frac{k_1 T^2}{2} - k_2 T + 1 \right)
    \\
    &\stackrel{!}{=} z^2 - 2 z_\mathrm{CL} \cdot z + z_\mathrm{CL}^2
    ,
\end{align*}
which we can solve for the controller gains $k_1$ and $k_2$, obtaining:
\begin{equation}
    k_1 = \frac{\left( 1 - z_\mathrm{CL} \right)^2}{T^2}
    \quad\text{and}\quad
    k_2 = \frac{4 - \left(1 + z_\mathrm{CL} \right)^2}{2 T}
    .
    \label{eqn:ADRC_DT_SS_Controller_K_n2}
\end{equation}

\subsection{Approximation for High Sampling Frequencies}
\label{sec:DT_Tuning_Approximation}

For the common case of sufficiently high sampling frequencies, we now want to provide a connection between discrete- and continuous-time controller tuning for ADRC, specifically for the practically relevant cases $n = 1$ and $n = 2$.

Let us revisit \refEq{eqn:ADRC_DT_zCL}, remembering the power series characterization of a real-valued exponential function.
This allows to express $z_\mathrm{CL}$ as:
\begin{align*}
    z_\mathrm{CL} &= \mathrm{e}^{-\omega_\mathrm{CL} T}
    \\
    &= 1 + \left( -\omega_\mathrm{CL} T \right) + \frac{\left( -\omega_\mathrm{CL} T \right)^2}{2!}
    + \frac{\left( -\omega_\mathrm{CL} T \right)^3}{3!} + \ldots
    ,
\end{align*}
which we can truncate after the second term if the sampling interval $T$ is small enough to warrant $\omega_\mathrm{CL} T \ll 1$.
Then we obtain the following approximation for $z_\mathrm{CL}$:
\begin{equation}
    z_\mathrm{CL} \approx 1 - \omega_\mathrm{CL} T
    .
    \label{eqn:ADRC_DT_zCL_Approximation}
\end{equation}

For the first-order case, putting \refEq{eqn:ADRC_DT_zCL_Approximation} in \refEq{eqn:ADRC_DT_SS_Controller_K_n1} gives:
\begin{equation}
    k_1 = \frac{1 - z_\mathrm{CL} }{ T }
    \approx \frac{1 - \left( 1 - \omega_\mathrm{CL} T \right) }{ T }
    = \omega_\mathrm{CL}
    ,
    \label{eqn:ADRC_DT_SS_Controller_K_n1_approx}
\end{equation}
which corresponds to the continuous-time tuning value obtained by \refEq{eqn:ADRC_CT_SS_Controller_K}.

In the same way, we obtain the following approximations for second-order ADRC:
\begin{gather}
    k_1 = \frac{\left( 1 - z_\mathrm{CL} \right)^2}{T^2}
    \approx \frac{\left( 1 - \left( 1 - \omega_\mathrm{CL} T \right) \right)^2}{T^2}
    = \omega_\mathrm{CL}^2
    ,
    \label{eqn:ADRC_DT_SS_Controller_K_n2_approx}
    \\
    \begin{aligned}
        k_2 = \frac{4 - \left(1 + z_\mathrm{CL} \right)^2}{2 T}
        &\approx \frac{4 - \left(1 + \left( 1 - \omega_\mathrm{CL} T \right) \right)^2}{2 T}
        \\
        &= \omega_\mathrm{CL} \cdot \left(2 - \frac{\omega_\mathrm{CL} T}{2} \right)
        \approx 2 \omega_\mathrm{CL}
        .
    \end{aligned}
    \notag
\end{gather}

We can now summarize the first contributions of this article:
\begin{itemize}
    \item
    It can be confirmed that the established quasi-continuous controller tuning can indeed be applied to discrete-time ADRC provided that the condition $\omega_\mathrm{CL} T \ll 1$ holds.

    \item
    The bandwidth parameterization tuning rules for the discrete-time controller derived in \refSec{sec:DT_Tuning_Bandwidth} are a ``backward compatible'' generalization of the existing practice, delivering correct results regardless of the choice of the sampling interval $T$.
\end{itemize}

\subsection{Comparison of Continuous- and Discrete-Time Tuning}
\label{sec:DT_Tuning_Comparison}

In \refSec{sec:ADRC_DT_Controller_Tuning} we mentioned that the established way of controller tuning within discrete-time ADRC is to simply use the continuous-time described in \refSec{sec:ADRC_CT_Controller_Tuning}.
The above-mentioned examples \cite{Miklosovic:2007,Zhang:2013,Stankovic:2016,Ahi:2018,Desai:2018,RamirezNeria:2020} obviously demonstrate as well that no problems are to be expected in most practical settings.
So when will the dedicated discrete-time tuning just introduced in \refSec{sec:DT_Tuning_Bandwidth} make a difference?

\begin{figure*}
    \centering%
    \subfloat[First-order ADRC]{%
        \includegraphics[scale=1]{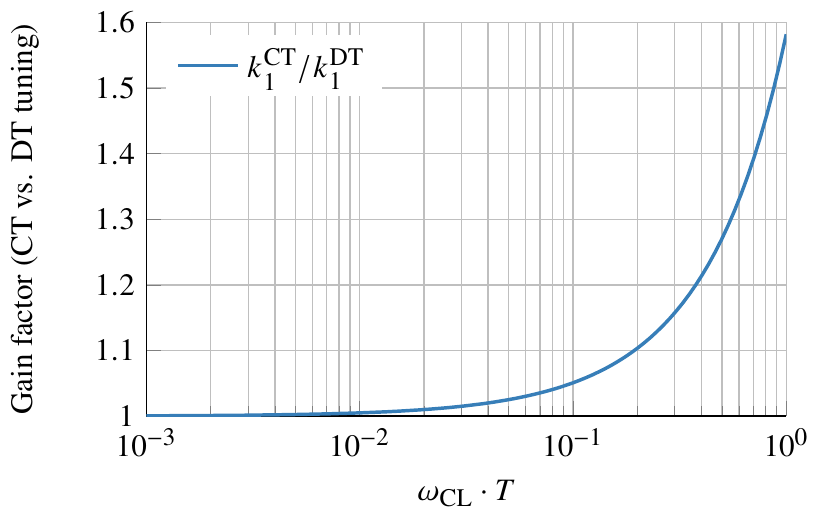}%
        \label{fig:CtrlTuning_DT_CT_Factor_1}
    }%
    \hfill%
    \subfloat[Second-order ADRC]{%
        \includegraphics[scale=1]{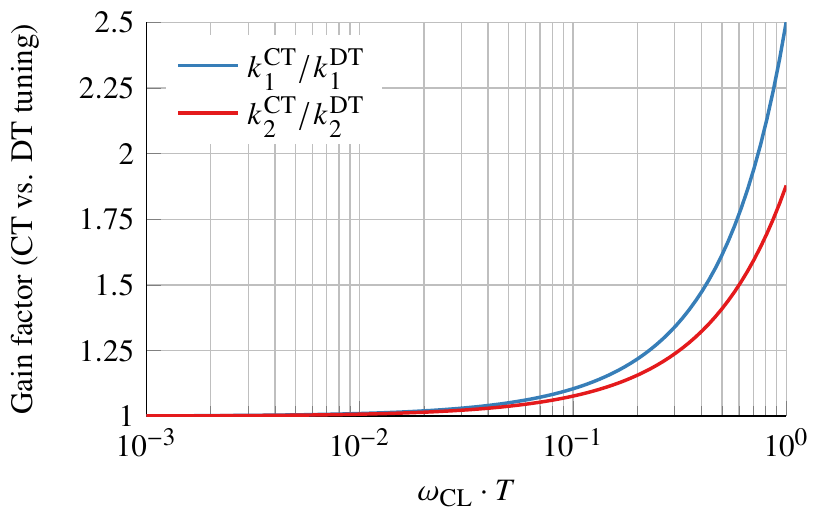}%
        \label{fig:CtrlTuning_DT_CT_Factor_2}
    }%
    \caption{%
        Relation of controller gains obtained by continuous-time design \refEq{eqn:ADRC_CT_SS_Controller_K_PolePlacement} vs.\ discrete-time design \refEq{eqn:ADRC_DT_SS_Controller_K_PolePlacement} depending on the relative sampling interval $\omega_\mathrm{CL} T$.
        As a practical rule of thumb, it can be said that below $\omega_\mathrm{CL} T \lessapprox 0.05$ both continuous- and discrete-time design become indistinguishable.
    }
    \label{fig:CtrlTuning_DT_CT_Factor}
\end{figure*}

Obviously the sampling interval $T$ must be small enough compared to the dynamics of the closed loop.
As already evident from \refSec{sec:DT_Tuning_Approximation}, we will therefore let our analysis be based on the relative factor $\omega_\mathrm{CL} T$.
To give a visual impression of the difference between continuous-time controller tuning as given by \refEq{eqn:ADRC_CT_SS_Controller_K_PolePlacement} and a discrete-time design obtained through \refEq{eqn:ADRC_DT_SS_Controller_K_PolePlacement}, the relative factor of continuous-time (CT) vs.\ discrete-time (DT) controller gains are shown in \refFig{fig:CtrlTuning_DT_CT_Factor} for first- and second-order ADRC.

For $n = 1$ (shown in \refFig{fig:CtrlTuning_DT_CT_Factor_1}), the relative factor is:
\begin{equation}
    \frac{ k_1^\mathrm{CT}(\omega_\mathrm{CL} T) }{ k_1^\mathrm{DT}(\omega_\mathrm{CL} T) }
    = \frac{ \omega_\mathrm{CL} }{ \displaystyle\frac{1 - z_\mathrm{CL} }{ T } }
    = \frac{ \omega_\mathrm{CL} T }{ 1 - \mathrm{e}^{-\omega_\mathrm{CL} T} }
    ,
    \label{eqn:CtrlTuning_DT_CT_Factor_1}
\end{equation}
while for $n = 2$ (shown in \refFig{fig:CtrlTuning_DT_CT_Factor_2}), the following two relations are obtained:
\begin{equation}
    \begin{aligned}
        &\frac{ k_1^\mathrm{CT}(\omega_\mathrm{CL} T) }{ k_1^\mathrm{DT}(\omega_\mathrm{CL} T) }
        = \frac{ \omega_\mathrm{CL}^2 }{ \displaystyle\frac{\left( 1 - z_\mathrm{CL} \right)^2}{T^2} }
        = \frac{ (\omega_\mathrm{CL} T)^2 }{ \left( 1 - \mathrm{e}^{-\omega_\mathrm{CL} T} \right)^2 }
        ,\\
        &\frac{ k_2^\mathrm{CT}(\omega_\mathrm{CL} T) }{ k_2^\mathrm{DT}(\omega_\mathrm{CL} T) }
        = \frac{ 2 \omega_\mathrm{CL} }{ \displaystyle\frac{4 - \left(1 + z_\mathrm{CL} \right)^2}{2 T} }
        = \frac{ 4 \omega_\mathrm{CL} T }{ 4 - \left( 1 + \mathrm{e}^{-\omega_\mathrm{CL} T} \right)^2 }
        .
    \end{aligned}
    \label{eqn:CtrlTuning_DT_CT_Factor_2}
\end{equation}

\begin{figure*}
    \includegraphics[scale=1]{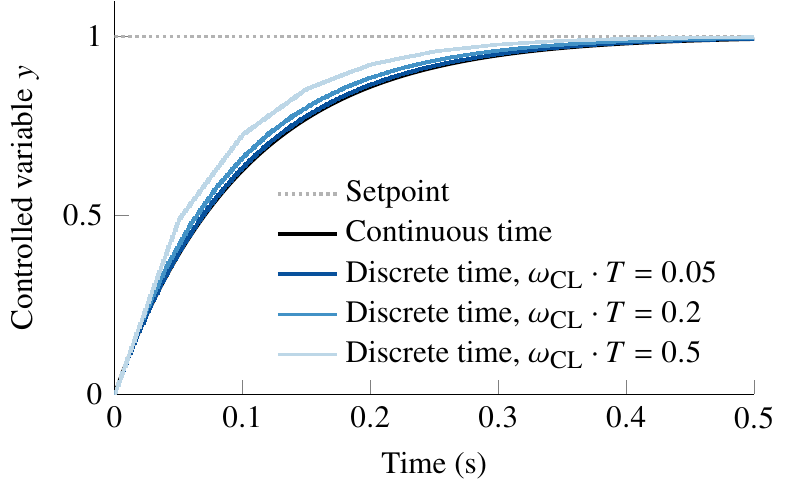}%
    \hfill%
    \includegraphics[scale=1]{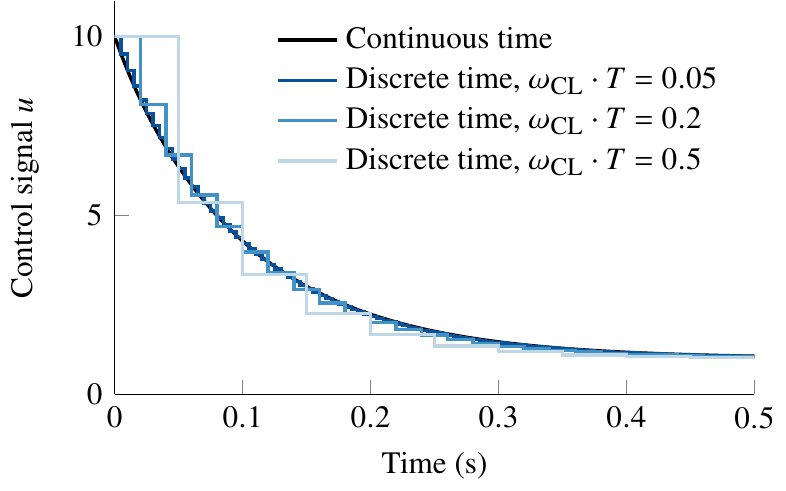}%
    \caption{%
        Comparison of closed-loop step response and control signal for a control loop with a normalized, continuous-time plant $P(s) = \frac{1}{s + 1}$ and first-order ADRC implementations tuned with $b_0 = 1$, $\omega_\mathrm{CL} = 10$, and $k_\mathrm{ESO} = 10$.
        For the discrete-time implementations with a ``traditional'' quasi-continuous controller design described in \refSec{sec:ADRC_CT_Controller_Tuning}, the actual bandwidth deviates from the true continuous-time case with increasing sampling intervals---the controller gains are too high.
        These simulations are idealized (no measurement noise, zero delay between measurement of $y$ and update of $u$) to focus the attention on the effects of increased sampling intervals.
    }
    \label{fig:CtrlTuning_ADRC1Compare_CT}
\end{figure*}

\begin{figure*}
    \includegraphics[scale=1]{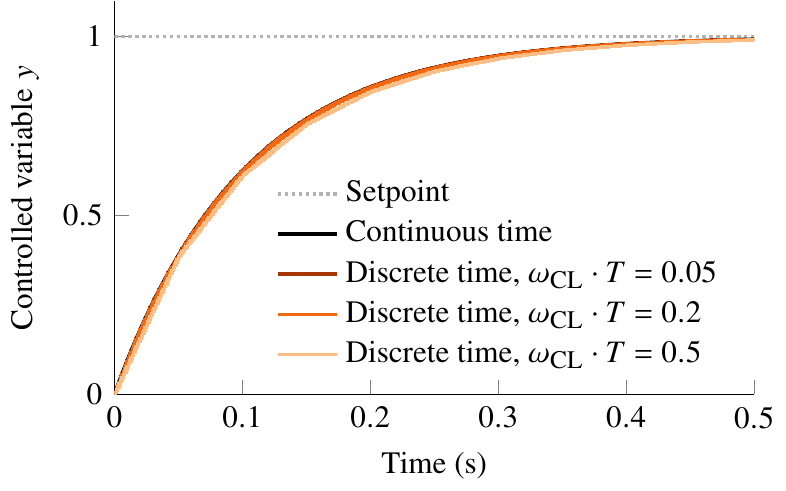}%
    \hfill%
    \includegraphics[scale=1]{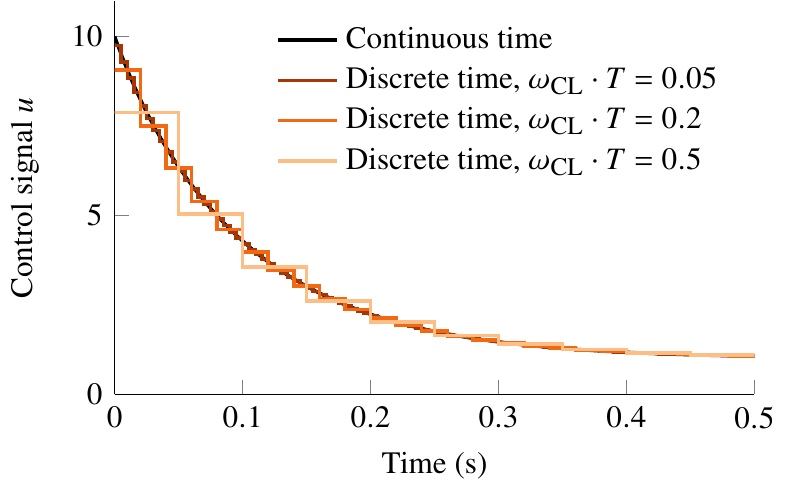}%
    \caption{%
        Comparison of \refFig{fig:CtrlTuning_ADRC1Compare_CT} repeated with the discrete-time controller design introduced in \refSec{sec:DT_Tuning_Bandwidth}, instead of the quasi-continuous design previously employed for discrete-time ADRC.
        Continuous- and discrete-time implementations now always exhibit similar closed-loop dynamics
    }
    \label{fig:CtrlTuning_ADRC1Compare_DT}
\end{figure*}

From the visualization of \refEq{eqn:CtrlTuning_DT_CT_Factor_1} and \refEq{eqn:CtrlTuning_DT_CT_Factor_2} in \refFig{fig:CtrlTuning_DT_CT_Factor} it becomes obvious that using a continuous-time design in a discrete-time ADRC implementation may lead to serious problems above $\omega_\mathrm{CL} T \gtrapprox 0.1$.
The controller gains obtained from continuous-time design are then overestimating the actually needed values, leading to a higher bandwidth than designed---and being closer to possible instability of the loop.

Conversely, a simple rule of thumb can be learned from \refFig{fig:CtrlTuning_DT_CT_Factor}.
For relative sampling intervals below $\omega_\mathrm{CL} T \lessapprox 0.05$, the closed-loop behaviors of a continuous- vs.\ discrete-time design will hardly be discernible.

Both findings are confirmed in a simple time-domain comparison of control loops with different discrete-time ADRC implementations using quasi-continuous and actual discrete-time controller designs in \refFig{fig:CtrlTuning_ADRC1Compare_CT} and \refFig{fig:CtrlTuning_ADRC1Compare_DT}.

As a final side note it shall be mentioned that for $\omega_\mathrm{CL} T \gtrapprox 0.1$ there will, of course, be other problems to tackle in the control loop besides overestimated controller gains. With such large sampling intervals, any delay unaccounted for will quickly lead to oscillations or even instability.

\section{Discrete-Time Transfer Function Implementations}
\label{sec:DT_TF}

In this section, we reiterate the derivation of two discrete-time transfer function implementations of ADRC introduced in \cite{Herbst:2021a} and \cite{Herbst:2021b}, with slight variations that lead to easier understanding and simpler tuning equations.
In contrast to previous work, however, we now give detailed coefficients for all transfer functions based on the discrete-time controller tuning rules developed in \refSec{sec:DT_Tuning_Bandwidth}.
This will also form the basis of introducing new discrete-time implementations of error-based ADRC in \refSec{sec:ErrorBased}.

\subsection{Transfer Function Implementation}
\label{sec:DT_TF_1}

The first step towards a discrete-time transfer function representation of the state-space implementation is made by $z$-transform of both controller \refEq{eqn:ADRC_DT_SS_Controller} and observer \refEq{eqn:ADRC_DT_SS_Observer}:
\begin{gather}
    u(z) = \frac{1}{b_0} \cdot \left( k_1 \cdot r(z) - \begin{pmatrix}
        \Vector{k}^\Transpose  &  1
    \end{pmatrix}
    \cdot \Vector{\hat{x}}(z) \right)
    \label{eqn:ADRC_DT_TF_ControlLaw_z}
    ,\\
    \Vector{\hat{x}}(z) = z^{-1} \cdot \Matrix{A}_\mathrm{ESO} \cdot \Vector{\hat{x}}(z) + z^{-1} \cdot \Vector{b}_\mathrm{ESO} \cdot u_\mathrm{lim}(z) + \Vector{l} \cdot y(z)
    .
    \label{eqn:ADRC_DT_TF_Observer_z}
\end{gather}

Putting \refEq{eqn:ADRC_DT_TF_ControlLaw_z} in \refEq{eqn:ADRC_DT_TF_Observer_z} leads to the closed-loop dynamics of the discrete-time observer.
To be able to do that, we sacrifice the separate feedback path of the limited control signal  $u_\mathrm{lim}$ back to the observer, as this feedback loop will now be ``convoluted'' in the closed-loop observer dynamics.
With $u(z) = u_\mathrm{lim}(z)$ one obtains:
\begin{equation}
    \Vector{\hat{x}}(z)
    =
    \Matrix{\Phi}_\mathrm{ESO} \cdot
    \left( z^{-1} \cdot \frac{k_1}{b_0} \cdot \Vector{b}_\mathrm{ESO} \cdot r(z) + \Vector{l} \cdot y(z) \right)
    .
    \label{eqn:ADRC_DT_TF_Observer_z_ClosedLoop}
\end{equation}

In \refEq{eqn:ADRC_DT_TF_Observer_z_ClosedLoop}, we introduced the matrix $\Matrix{\Phi}_\mathrm{ESO}$ as a shorthand notation:
\begin{equation}
    \Matrix{\Phi}_\mathrm{ESO}
    =
    \left(
    \Matrix{I} - z^{-1} \cdot \left( \Matrix{A}_\mathrm{ESO} - \frac{1}{b_0} \cdot  \Vector{b}_\mathrm{ESO} \cdot \begin{pmatrix} \Vector{k}^\Transpose & 1 \end{pmatrix} \right)
    \right)^{-1}
    .
    \label{eqn:ADRC_DT_TF_Observer_Phi}
\end{equation}

Transfer functions will relate the reference signal and the measured plant output to the updated controller output.
The vector of estimated state variables $\Vector{\hat{x}}(z)$ therefore needs to be eliminated.
That can be achieved by by putting \refEq{eqn:ADRC_DT_TF_Observer_z_ClosedLoop} back into \refEq{eqn:ADRC_DT_TF_ControlLaw_z}.
The control law can now be written as:
\begin{align}
    u(z)
    &=
    \frac{k_1}{b_0} \cdot \left( 1 - z^{-1} \cdot \begin{pmatrix} \Vector{k}^\Transpose & 1 \end{pmatrix} \cdot \Matrix{\Phi}_\mathrm{ESO} \cdot \frac{1}{b_0} \cdot \Vector{b}_\mathrm{ESO} \right) \cdot r(z)
    \notag
    \\
    &\phantom{=}\ - \left( \frac{1}{b_0} \cdot \begin{pmatrix} \Vector{k}^\Transpose & 1 \end{pmatrix} \cdot \Matrix{\Phi}_\mathrm{ESO} \cdot \Vector{l} \right) \cdot y(z)
    .
    \label{eqn:ADRC_DT_TF_ControlLaw_Discrete_ry}
\end{align}

\begin{figure}
    \includegraphics[width=\linewidth]{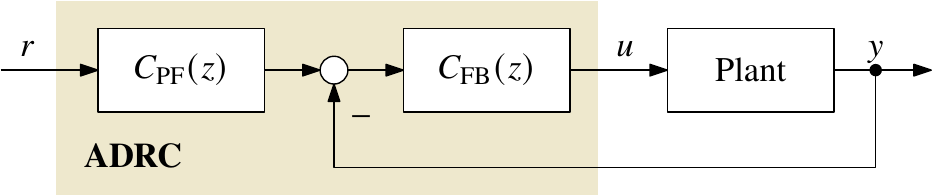}
    \caption{Discrete-time transfer function implementation of ADRC}
    \label{fig:ADRC_DT_TF}
\end{figure}

\refEqBegin{eqn:ADRC_DT_TF_ControlLaw_Discrete_ry} obviously represents a two-degrees-of-freedom control structure, albeit not a typical (or desirable) one.
A discrete-time controller structure often encountered in practice is shown in \refFig{fig:ADRC_DT_TF} and consists of a feedback controller $C_\mathrm{FB}(z)$ and a prefilter $C_\mathrm{PF}(z)$ as follows:
\begin{equation}
    u(z) = C_\mathrm{FB}(z) \cdot \left[ C_\mathrm{PF}(z) \cdot r(z) - y(z) \right]
    .
    \label{eqn:ADRC_DT_TF_Controller}
\end{equation}

This structure is also shown in \refFig{fig:ADRC_DT_TF}.
We can transform \refEq{eqn:ADRC_DT_TF_ControlLaw_Discrete_ry} into that form by factoring out the feedback transfer function.
Now we can read off the transfer functions of both feedback controller $C_\mathrm{FB}(z)$ and prefilter $C_\mathrm{PF}(z)$ as follows:
\begin{align}
    C_\mathrm{FB}(z)
    &= \frac{1}{b_0} \cdot \begin{pmatrix} \Vector{k}^\Transpose & 1 \end{pmatrix} \cdot \Matrix{\Phi}_\mathrm{ESO} \cdot \Vector{l},
    \label{eqn:ADRC_DT_TF_CFB_General}
    \\
    C_\mathrm{PF}(z)
    &= \frac{ \frac{k_1}{b_0} \cdot \left( 1 - z^{-1} \cdot \begin{pmatrix} \Vector{k}^\Transpose & 1 \end{pmatrix} \cdot \Matrix{\Phi}_\mathrm{ESO} \cdot \frac{1}{b_0} \cdot \Vector{b}_\mathrm{ESO} \right) }{ \frac{1}{b_0} \cdot \begin{pmatrix} \Vector{k}^\Transpose & 1 \end{pmatrix} \cdot \Matrix{\Phi}_\mathrm{ESO} \cdot \Vector{l} }
    \label{eqn:ADRC_DT_TF_CPF_General}
    .
\end{align}

As its counterpart in the continuous time domain, the discrete-time feedback controller transfer function $C_\mathrm{FB}(z)$ includes an integrator\textemdash otherwise the control loop would not be able to reject constant disturbances and follow constant reference signal values with zero steady-state error, obviously.
In \cite{Herbst:2021a} it was put forward to factor this pole at $z = 1$ out of $C_\mathrm{FB}(z)$:
\begin{equation*}
    C_\mathrm{FB}(z) =
    \Delta C_\mathrm{FB}(z) \cdot \frac{1}{1 - z^{-1}}
    .
\end{equation*}

Implementing $C_\mathrm{FB}(z)$ as a series connection of an accumulator and a discrete-time filter comes with the following advantages:
\begin{itemize}
    \item
    One dependent coefficient of the denominator polynomial is removed, which means there is one coefficient less to compute and to store in an implementation.
    This also saves one multiplication at runtime.

    \item
    The $\Delta C_\mathrm{FB}(z)$ part of the feedback controller now exhibits the \emph{incremental} \cite{Peng:1996} or \emph{velocity} \cite{Astrom:2006} form of a digital controller.
    Implementing the accumulator with clamping therefore provides an easy but quite effective protection against integrator windup.

\end{itemize}

This is also the reason why \refEq{eqn:ADRC_DT_TF_ControlLaw_Discrete_ry} had to be brought into the form \refEq{eqn:ADRC_DT_TF_Controller}---otherwise there would be a second integrator, namely in the transfer function from $r$ to $u$.
For the overall behavior, these two structures are equivalent in a mathematical sense, but having two integrators (one between $r$ and $u$ and one between $y$ and $u$) would result in two integrator states running away to infinity for any nonzero reference $r$.

Putting everything together and applying discrete-time observer and controller tuning as described in \refSec{sec:ADRC_DT_Observer_Tuning} and \refSec{sec:DT_Tuning_Bandwidth}, one obtains the following two transfer functions $C_{\mathrm{FB}}(z)$ and $C_{\mathrm{PF}}(z)$:
\begin{align}
    C_{\mathrm{FB}}(z) &= \frac
    { \displaystyle\sum_{i = 0}^n \beta_i z^{-i} }
    { 1 + \displaystyle\sum_{i = 1}^n \alpha_i z^{-i} }
    \cdot \frac{1}{1 - z^{-1}}
    \label{eqn:ADRC_DT_TF_CFB}
    ,\\
    C_{\mathrm{PF}}(z) &= \frac
    { \displaystyle\frac{1}{\beta_0} \displaystyle\sum_{i = 0}^{n+1} \gamma_i z^{-i} }
    { 1 + \displaystyle\frac{1}{\beta_0} \sum_{i = 1}^n \beta_i z^{-i} }
    .
    \label{eqn:ADRC_DT_TF_CPF}
\end{align}

For the first- and second-order case of ADRC, which are particularly relevant in practice, we provide a detailed set of equations for the $\alpha$, $\beta$, $\gamma$ coefficients based on the continuous-time tuning parameters $\omega_\mathrm{CL}$ (desired closed-loop bandwidth) and $k_\mathrm{ESO}$ (relative observer bandwidth factor) in \refTable{table:DT_TF_Coefficients_1} and \refTable{table:DT_TF_Coefficients_2}, respectively.
These results are---in the sense of \refSec{sec:DT_Tuning_Comparison}---a more accurate generalization of the coefficients previously reported in \cite{Herbst:2021a}, which used a quasi-continuous design for the controller gains.
In \refEq{eqn:ADRC_DT_TF_CPF} we factored out $\frac{1}{\beta_0}$, which allows to reuse the $\beta$ coefficients for $C_{\mathrm{PF}}(z)$ and give more compact tuning equations for the $\gamma$ coefficients.

\begin{table}
    \caption{%
        Discrete-time transfer function implementation of first-order ADRC:
        coefficients of the transfer functions $C_\mathrm{FB}(z)$ and $C_\mathrm{PF}(z)$
        from \refEq{eqn:ADRC_DT_TF_CFB} and \refEq{eqn:ADRC_DT_TF_CPF},
        obtained by discrete-time pole placement and bandwidth parameterization with
        $z_\mathrm{CL} = \mathrm{e}^{-\omega_\mathrm{CL} T}$ (closed loop) and
        $z_\mathrm{ESO} = \mathrm{e}^{-k_\mathrm{ESO} \omega_\mathrm{CL} T}$ (observer)
    }
    \label{table:DT_TF_Coefficients_1}
    \begin{tabular*}{\linewidth}{@{\extracolsep\fill}rl@{\extracolsep\fill}}
        \toprule
        \textbf{Coefficient}  &  \textbf{Value for first-order ADRC}  \\
        \midrule
        \\[-0.9em]
        $\alpha_1$
        &   $-z_\mathrm{CL} z_\mathrm{ESO}^2$
        \\[1.0em]
        $\beta_0$
        &   $\displaystyle\frac{1}{b_0 T} \cdot \left[
                z_\mathrm{CL} z_\mathrm{ESO}^2 - 2 z_\mathrm{ESO} - z_\mathrm{CL} + 2
            \right]$
        \\[1.5em]
        $\beta_1$
        &   $\displaystyle\frac{1}{b_0 T} \cdot \left[
                2 z_\mathrm{CL} z_\mathrm{ESO} - 2 z_\mathrm{CL} z_\mathrm{ESO}^2 + z_\mathrm{ESO}^2 - 1
            \right]$
        \\[1.5em]
        $\gamma_0$
        &   $\displaystyle\frac{1 - z_\mathrm{CL}}{b_0 T}$
        \\[1.5em]
        $\gamma_1$
        &   $\displaystyle\frac{-2 z_\mathrm{ESO} \cdot (1 - z_\mathrm{CL})}{b_0 T}$
        \\[1.5em]
        $\gamma_2$
        &   $\displaystyle\frac{z_\mathrm{ESO}^2 \cdot (1 - z_\mathrm{CL})}{b_0 T}$
        \\[1.5em]
        \bottomrule
    \end{tabular*}
\end{table}

\begin{table*}
    \caption{%
        Discrete-time transfer function implementation of second-order ADRC:
        coefficients of the transfer functions $C_\mathrm{FB}(z)$ and $C_\mathrm{PF}(z)$
        from \refEq{eqn:ADRC_DT_TF_CFB} and \refEq{eqn:ADRC_DT_TF_CPF},
        obtained by discrete-time pole placement and bandwidth parameterization with
        $z_\mathrm{CL} = \mathrm{e}^{-\omega_\mathrm{CL} T}$ (closed loop) and
        $z_\mathrm{ESO} = \mathrm{e}^{-k_\mathrm{ESO} \omega_\mathrm{CL} T}$ (observer).
        Underlying system and tuning parameters are $b_0$ (plant gain), $T$ (sampling interval), $\omega_\mathrm{CL}$ (closed-loop bandwidth), $k_\mathrm{ESO}$ (observer bandwidth factor)
    }
    \label{table:DT_TF_Coefficients_2}
    \begin{tabular*}{\textwidth}{@{\extracolsep\fill}rl@{\extracolsep\fill}}
        \toprule
        \textbf{Coefficient}  &  \textbf{Value for second-order ADRC}  \\
        \midrule
        \\[-0.9em]
        $\alpha_1$
        &   $\displaystyle -\frac{1}{8} \cdot (1 + z_\mathrm{CL})^2 \cdot (1 + z_\mathrm{ESO})^3 + z_\mathrm{CL}^2 z_\mathrm{ESO}^3 + 1$
        \\[1.0em]
        $\alpha_2$
        &   $z_\mathrm{CL}^2 z_\mathrm{ESO}^3$
        \\[1.5em]
        $\beta_0$
        &   $\displaystyle\frac{1}{b_0 T^2} \cdot \left[
                \frac{1}{4} \cdot (1 + z_\mathrm{CL})^2 \cdot (1 + z_\mathrm{ESO})^3
                - 2 \cdot \left( z_\mathrm{CL}^2  z_\mathrm{ESO}^3 + 2 z_\mathrm{CL} + 3 z_\mathrm{ESO} - 2 \right)
            \right]$
        \\[1.0em]
        $\beta_1$
        &   $\displaystyle\frac{1}{b_0 T^2} \cdot \left[
                -(1 + z_\mathrm{CL})^2 \cdot (1 + z_\mathrm{ESO})^3 + 2 \cdot (1 + z_\mathrm{CL})^2
                + 6 \cdot \left( z_\mathrm{CL}^2 z_\mathrm{ESO}^3 + 2 z_\mathrm{CL} z_\mathrm{ESO} + z_\mathrm{ESO}^2 + z_\mathrm{ESO} - 1 \right)
            \right]$
        \\[1.0em]
        $\beta_2$
        &   $\displaystyle\frac{1}{b_0 T^2} \cdot \left[
                -\frac{1}{4} \cdot (1 + z_\mathrm{CL})^2 \cdot (1 + z_\mathrm{ESO})^3
                + 2 \cdot \left( -2 z_\mathrm{CL}^2 z_\mathrm{ESO}^3 + 3 z_\mathrm{CL}^2 z_\mathrm{ESO}^2 + 2 z_\mathrm{CL} z_\mathrm{ESO}^3 + 1 \right)
            \right]$
        \\[1.5em]
        $\gamma_0$
        &   $\displaystyle\frac{(1 - z_\mathrm{CL})^2}{b_0 T^2}$
        \\[1.5em]
        $\gamma_1$
        &   $\displaystyle\frac{-3 z_\mathrm{ESO} \cdot (1 - z_\mathrm{CL})^2}{b_0 T^2}$
        \\[1.5em]
        $\gamma_2$
        &   $\displaystyle\frac{3 z_\mathrm{ESO}^2 \cdot (1 - z_\mathrm{CL})^2}{b_0 T^2}$
        \\[1.5em]
        $\gamma_3$
        &   $\displaystyle\frac{-z_\mathrm{ESO}^3 \cdot (1 - z_\mathrm{CL})^2}{b_0 T^2}$
        \\[1.5em]
        \bottomrule
    \end{tabular*}
\end{table*}

Note that this discrete-time transfer function implementation does not fully replicate the behavior of its state-space ancestor.
On the other hand, transfer functions can be implemented more efficiently, as their number of coefficients grows only linearly with the order of the controller $n$, as opposed to quadratic growth in the state-space case.
The price for obtaining a very familiar structure with prefilter and feedback controller was to eliminate the separate feedback path of the limited controller output $u_\mathrm{lim}$.
This results in loss of the ``out-of-the-box'' windup protection coming with ADRC---a loss than can be partially compensated by implementation of a clamped integrator as part of $C_{\mathrm{FB}}(z)$.
Luckily, it is not always an either/or situation.
In the second transfer function approach, that we will introduce in \refSec{sec:DT_TF_2}, state-space features will be combined with transfer function efficiency.

\subsection{Dual-Feedback Transfer Function Implementation}
\label{sec:DT_TF_2}

A different transfer function implementation that preserves the feedback path of the limited control signal value $u_\mathrm{lim}(z)$ was introduced in \cite{Herbst:2021b}.
The starting point of \refSec{sec:DT_TF_1} is being used here, as well: the control law \refEq{eqn:ADRC_DT_TF_ControlLaw_z} and observer \refEq{eqn:ADRC_DT_TF_Observer_z} in $z$-domain.
The latter is being solved for $\Vector{\hat{x}}(z)$:
\begin{equation*}
    \Vector{\hat{x}}(z) =
    \left( \Matrix{I} - z^{-1} \Matrix{A}_\mathrm{ESO} \right)^{-1}
    \cdot \left( z^{-1} \cdot \Vector{b}_\mathrm{ESO} \cdot u_\mathrm{lim}(z) + \Vector{l} \cdot y(z) \right)
    .
    \label{eqn:ADRC_DT_FBTF_Derivation_Observer2}
\end{equation*}

\begin{figure}
    \includegraphics[width=\linewidth]{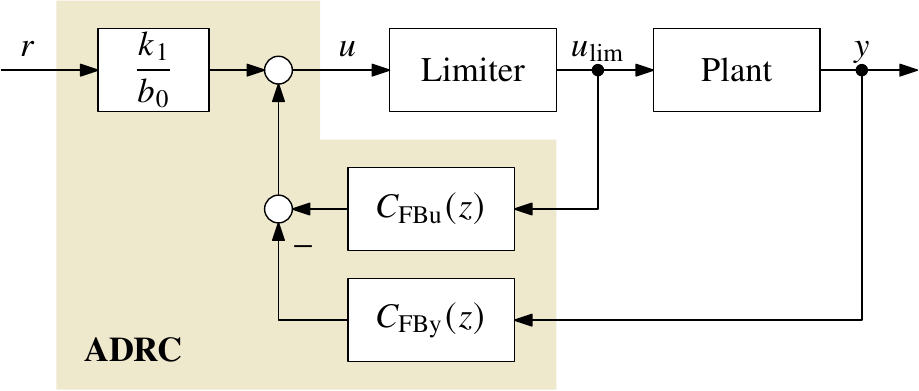}
    \caption{%
        Control loop with discrete-time ADRC, dual-feedback implementation using two transfer functions $C_\mathrm{FBy}(z)$ and $C_\mathrm{FBu}(z)$ given in \refEq{eqn:ADRC_DT_FBTF_FBy} and \refEq{eqn:ADRC_DT_FBTF_FBu}
    }
    \label{fig:ADRC_DT_FBTF}
\end{figure}

Putting this equation back in \refEq{eqn:ADRC_DT_TF_ControlLaw_z} yields a control law depicted in \refFig{fig:ADRC_DT_FBTF}.
It features three inputs $r$, $y$, and $u_\mathrm{lim}$, as in the state-space case:
\begin{align}
    u(z) & = \frac{1}{b_0} \cdot \Big(
    k_1 \cdot r(z) - \begin{pmatrix} \Vector{k}^\Transpose & 1 \end{pmatrix}
    \cdot \left( \Matrix{I} - z^{-1} \Matrix{A}_\mathrm{ESO} \right)^{-1}
    \notag\\
    & \quad\quad\quad\quad\quad \cdot \left(
        z^{-1} \cdot \Vector{b}_\mathrm{ESO} \cdot u_\mathrm{lim}(z) + \Vector{l} \cdot y(z)
    \right) \Big)
    \notag\\
    &= \frac{k_1}{b_0} \cdot r(z)
    - C_\mathrm{FBy}(z) \cdot y(z)
    + C_\mathrm{FBu}(z) \cdot u_\mathrm{lim}(z)
    .
    \label{eqn:ADRC_DT_FBTF_Controller}
\end{align}

The two feedback transfer functions $C_\mathrm{FBy}(z)$ and $C_\mathrm{FBu}(z)$ in \refEq{eqn:ADRC_DT_FBTF_Controller} share the same denominator and can be expressed as given in \refEq{eqn:ADRC_DT_FBTF_FBy} and \refEq{eqn:ADRC_DT_FBTF_FBu}.

\begin{align}
    C_\mathrm{FBy}(z)
    &= \frac{1}{b_0} \cdot \begin{pmatrix}
        \Vector{k}^\Transpose  &  1
    \end{pmatrix}
    \cdot \left( \Matrix{I} - z^{-1} \Matrix{A}_\mathrm{ESO} \right)^{-1}
    \cdot \Vector{l}
    \notag\\
    &= \frac{
        \displaystyle\sum_{i = 0}^{n} \beta_i z^{-i}
    }{
        1 + \displaystyle\sum_{i = 1}^{n+1} \alpha_i z^{-i}
    }
    \label{eqn:ADRC_DT_FBTF_FBy}
\end{align}

\begin{align}
    C_\mathrm{FBu}(z)
    &= -\frac{z^{-1}}{b_0} \cdot \begin{pmatrix}
        \Vector{k}^\Transpose  &  1
    \end{pmatrix}
    \cdot \left( \Matrix{I} - z^{-1} \Matrix{A}_\mathrm{ESO} \right)^{-1}
    \cdot \Vector{b}_\mathrm{ESO}
    \notag\\
    &= z^{-1} \cdot \frac{
        \displaystyle\sum_{i = 0}^{n} \gamma_i z^{-i }
    }{
        1 + \displaystyle\sum_{i = 1}^{n+1} \alpha_i z^{-i}
    }
    \label{eqn:ADRC_DT_FBTF_FBu}
\end{align}

The transfer function coefficients are computed by inserting properly tuned controller gains $\Vector{k}^\Transpose$ and observer gains $\Vector{l}$ in \refEq{eqn:ADRC_DT_FBTF_Controller} and solving for $\alpha$, $\beta$, and $\gamma$.
Reference \cite{Herbst:2021b} gave tuning equations for these coefficients using bandwidth parameterization with the quasi-continuous controller design of \refSec{sec:ADRC_CT_Controller_Tuning}.
In this paper, we provide these coefficients for first- and second-order ADRC incorporating the discrete-time controller tuning of \refSec{sec:DT_Tuning_Bandwidth}, now providing accurate results also for larger sampling intervals.
They can be found in \refTable{table:DT_FBTF_Coefficients_1} and \refTable{table:DT_FBTF_Coefficients_2}.
Note that the $\beta$ coefficients are the same as in \refSec{sec:DT_TF_1}, i.\,e.\ the numerator polynomial of $C_\mathrm{FBy}(z)$ in \refEq{eqn:ADRC_DT_FBTF_FBy} and $C_\mathrm{FB}(z)$ in \refEq{eqn:ADRC_DT_TF_CFB} are identical.

\begin{table}
    \caption{%
        Discrete-time dual-feedback transfer function implementation:
        coefficients for first-order ADRC,
        obtained by discrete-time pole placement and bandwidth parameterization with
        $z_\mathrm{CL} = \mathrm{e}^{-\omega_\mathrm{CL} T}$ (closed loop) and
        $z_\mathrm{ESO} = \mathrm{e}^{-k_\mathrm{ESO} \omega_\mathrm{CL} T}$ (observer)
    }
    \label{table:DT_FBTF_Coefficients_1}
    \begin{tabular*}{\linewidth}{@{\extracolsep\fill}rl@{\extracolsep\fill}}
        \toprule
        \textbf{Coefficient}  &  \textbf{Value for first-order ADRC}  \\
        \midrule
        \\[-0.9em]
        $\alpha_1$
        &   $-2 z_\mathrm{ESO}$
        \\[1.0em]
        $\alpha_2$
        &   $z_\mathrm{ESO}^2$
        \\[1.0em]
        $\beta_0$
        &   $\displaystyle\frac{1}{b_0 T} \cdot \left[
                z_\mathrm{CL} z_\mathrm{ESO}^2 - 2 z_\mathrm{ESO} - z_\mathrm{CL} + 2
            \right]$
        \\[1.5em]
        $\beta_1$
        &   $\displaystyle\frac{1}{b_0 T} \cdot \left[
                2 z_\mathrm{CL} z_\mathrm{ESO} - 2 z_\mathrm{CL} z_\mathrm{ESO}^2 + z_\mathrm{ESO}^2 - 1
            \right]$
        \\[1.5em]
        $\gamma_0$
        &   $z_\mathrm{CL} z_\mathrm{ESO}^2 - 2 z_\mathrm{ESO} + 1$
        \\[1.5em]
        $\gamma_1$
        &   $z_\mathrm{ESO}^2 - z_\mathrm{CL} z_\mathrm{ESO}^2$
        \\[1.5em]
        $\displaystyle\frac{k_1}{b_0}$
        &   $\displaystyle\frac{1 - z_\mathrm{CL}}{b_0 T}$
        \\[1.0em]
        \bottomrule
    \end{tabular*}
\end{table}

\begin{table*}
    \caption{%
        Discrete-time dual-feedback transfer function implementation:
        coefficients for second-order ADRC,
        obtained by discrete-time pole placement and bandwidth parameterization with
        $z_\mathrm{CL} = \mathrm{e}^{-\omega_\mathrm{CL} T}$ (closed loop) and
        $z_\mathrm{ESO} = \mathrm{e}^{-k_\mathrm{ESO} \omega_\mathrm{CL} T}$ (observer)
    }
    \label{table:DT_FBTF_Coefficients_2}
    \begin{tabular*}{\textwidth}{@{\extracolsep\fill}rl@{\extracolsep\fill}}
        \toprule
        \textbf{Coefficient}  &  \textbf{Value for second-order ADRC}  \\
        \midrule
        \\[-0.9em]
        $\alpha_1$
        &   $-3 z_\mathrm{ESO}$
        \\[1.0em]
        $\alpha_2$
        &   $3 z_\mathrm{ESO}^2$
        \\[1.0em]
        $\alpha_3$
        &   $-z_\mathrm{ESO}^3$
        \\[1.5em]
        $\beta_0$
        &   $\displaystyle\frac{1}{b_0 T^2} \cdot \left[
                \frac{1}{4} \cdot (1 + z_\mathrm{CL})^2 \cdot (1 + z_\mathrm{ESO})^3
                - 2 \cdot \left( z_\mathrm{CL}^2  z_\mathrm{ESO}^3 + 2 z_\mathrm{CL} + 3 z_\mathrm{ESO} - 2 \right)
            \right]$
        \\[1.5em]
        $\beta_1$
        &   $\displaystyle\frac{1}{b_0 T^2} \cdot \left[
                -(1 + z_\mathrm{CL})^2 \cdot (1 + z_\mathrm{ESO})^3 + 2 \cdot (1 + z_\mathrm{CL})^2
                + 6 \cdot \left( z_\mathrm{CL}^2 z_\mathrm{ESO}^3 + 2 z_\mathrm{CL} z_\mathrm{ESO} + z_\mathrm{ESO}^2 + z_\mathrm{ESO} - 1 \right)
            \right]$
        \\[1.5em]
        $\beta_2$
        &   $\displaystyle\frac{1}{b_0 T^2} \cdot \left[
                -\frac{1}{4} \cdot (1 + z_\mathrm{CL})^2 \cdot (1 + z_\mathrm{ESO})^3
                + 2 \cdot \left( -2 z_\mathrm{CL}^2 z_\mathrm{ESO}^3 + 3 z_\mathrm{CL}^2 z_\mathrm{ESO}^2 + 2 z_\mathrm{CL} z_\mathrm{ESO}^3 + 1 \right)
            \right]$
        \\[1.5em]
        $\gamma_0$
        &   $\displaystyle\frac{1}{8} \cdot (1 + z_\mathrm{CL})^2 \cdot (1 + z_\mathrm{ESO})^3 - z_\mathrm{ESO} \cdot (z_\mathrm{CL}^2 z_\mathrm{ESO}^2 + 3)$
        \\[1.0em]
        $\gamma_1$
        &   $\displaystyle-\frac{1}{8} \cdot (1 + z_\mathrm{CL})^2 \cdot (1 + z_\mathrm{ESO})^3 + 3 z_\mathrm{ESO}^2 + 1$
        \\[1.0em]
        $\gamma_2$
        &   $z_\mathrm{ESO}^3 \cdot \left( z_\mathrm{CL}^2 - 1 \right)$
        \\[1.5em]
        $\displaystyle\frac{k_1}{b_0}$
        &   $\displaystyle\frac{\left( 1 - z_\mathrm{CL} \right)^2}{b_0 T^2}$
        \\[1.0em]
        \bottomrule
    \end{tabular*}
\end{table*}

A very efficient implementation was proposed in \cite{Herbst:2021b}, exploiting the common denominator of $C_\mathrm{FBy}(z)$ and $C_\mathrm{FBu}(z)$ and choosing a transposed `direct form II' realization.
This reduces the number of required storage variables during runtime to the minimum, and uses even less algebraic operations than the transfer function representation of \refSec{sec:DT_TF_1}---while keeping the flexibility and behavior of the state-space implementation regarding user-defined control signal limitation and windup protection.
For the first- and second-order case, the resulting controller structure is shown in \refFig{fig:ADRC_DT_FBTF_1} and \refFig{fig:ADRC_DT_FBTF_2}.

\section{Error-Based ADRC}
\label{sec:ErrorBased}

Error-based ADRC is a variant presented in generalized form in \cite{Madonski:2019}.
It obtained its name from feeding the control error $e = r - y$ to the observer---instead of the plant output $y$ in the ``classical'' case like the one shown in \refSec{sec:ADRC_CT}.
To better differentiate the two approaches, we will accordingly denote the latter as ``output-based'' ADRC in this section.

The relation between output- and error-based ADRC was firmly established for the continuous-time case in \cite{Madonski:2023a}.
Building on that, we want to extend the connections between output- and error-based ADRC to the discrete time domain in this article.
The only discrete-time implementation of error-based ADRC so far in \cite{Madonski:2019} was based on a bilinear transformation from its continuous-time transfer function representation.
As we will see in this section, we can fully build on existing discrete-time ADRC structures instead.
In doing so, we will derive new, equivalent variants of discrete-time error-based ADRC that match the three implementations covered in this paper: state-space, transfer function, and dual-feedback transfer function form.
As an extra gain, this will extend error-based ADRC with the control signal limitation and anti-windup abilities of output-based ADRC.

\subsection{Continuous-Time Error-Based ADRC}
\label{sec:ErrorBased_CT}

\begin{figure}
    \includegraphics[width=\linewidth]{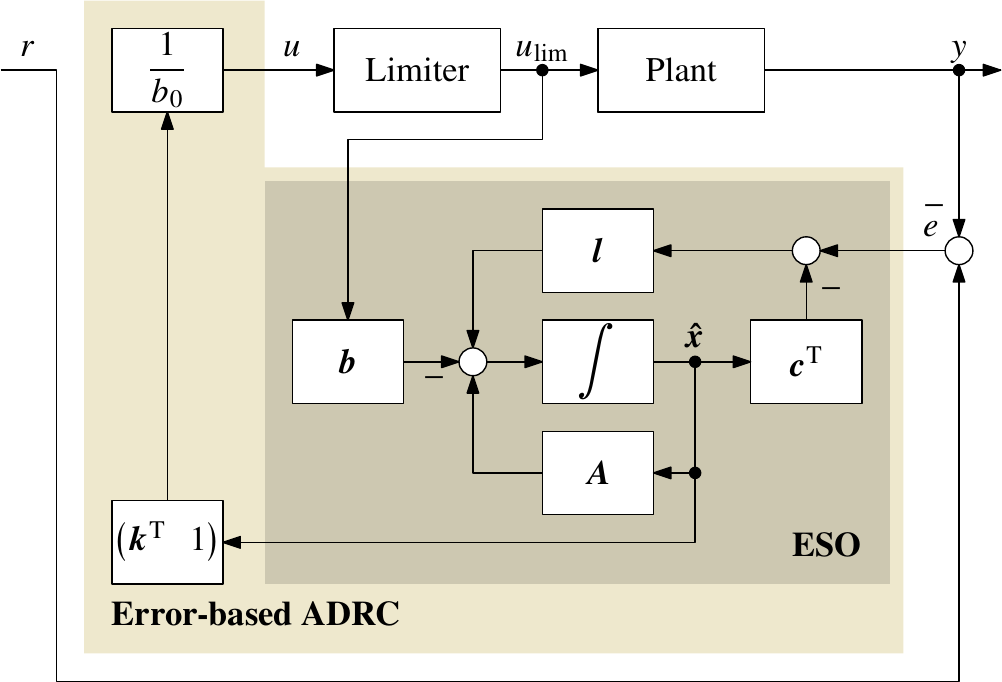}
    \caption{Structure of a control loop with error-based ADRC (continuous-time state-space version) and a user-defined control signal limiter}
    \label{fig:ADRC_CTe_SS}
\end{figure}

For an in-depth introduction to error-based ADRC, we refer to \cite{Madonski:2019}.
A continuous-time block diagram of error-based ADRC is shown in \refFig{fig:ADRC_CTe_SS}.
In contrast to \cite{Madonski:2019}, we explicitly are feeding the limited control signal $u_\mathrm{lim}(t)$ into to observer, to provide easy windup protection as in the ouput-based case in \refSec{sec:ADRC_CT_ControlLaw}.
The associated control law and observer dynamics are:
\begin{gather}
    u(t) = \frac{1}{b_0} \cdot \begin{pmatrix} \Vector{k}^\Transpose & 1 \end{pmatrix}
    \cdot \Vector{\hat{x}}(t)
    ,
    \label{eqn:ADRC_CTe_SS_Controller}
    \\
    \Vector{\dot{\hat{x}}}(t) = \Matrix{A} \cdot \Vector{\hat{x}}(t) - \Vector{b} \cdot u_\mathrm{lim}(t) + \Vector{l} \cdot \left( e(t) - \Vector{c}^\Transpose \cdot \Vector{\hat{x}}(t) \right)
    .
    \label{eqn:ADRC_CTe_SS_Observer}
\end{gather}

Not only the structure of these equations and the block diagram are similar to output-based ADRC:
In the observer \refEq{eqn:ADRC_CTe_SS_Observer}, $\Matrix{A}$, $\Vector{b}$, and  $\Vector{c}^\Transpose$ are exactly the same as in \refEq{eqn:ADRC_CT_SS_Observer}.
This applies to the controller gains $\Vector{k}^\Transpose$ and observer gains $\Vector{l}$, as well, which means that all existing tuning methods can be used here without modification.

\subsection{Discrete-Time State-Space Implementation}
\label{sec:ErrorBased_DT}

To derive the discrete-time state-space version of error-based ADRC, we can largely build on the results from output-based ADRC in \refSec{sec:ADRC_DT_ControlLaw}.
The control law \refEq{eqn:ADRC_CTe_SS_Controller} is a static state feedback, and hence trivial to translate to the discrete time domain:
\begin{equation}
    u(k) = \frac{1}{b_0} \cdot \begin{pmatrix} \Vector{k}^\Transpose & 1 \end{pmatrix}
    \cdot \Vector{\hat{x}}(k)
    .
    \label{eqn:ADRC_DTe_SS_Controller}
\end{equation}

We only have to derive a discretized observer from \refEq{eqn:ADRC_CTe_SS_Observer}.
By visual inspection of \refFig{fig:ADRC_CT_SS} and \refFig{fig:ADRC_CTe_SS}, only two differences between the observers can be found:
\begin{itemize}
    \item
    For error-based ADRC, the input $e(t)$ is being used instead of $y(t)$, and
    \item
    the limited controller output $u_\mathrm{lim}(t)$ enters the observer with a negative sign.
\end{itemize}

\begin{figure}
    \includegraphics[width=\linewidth]{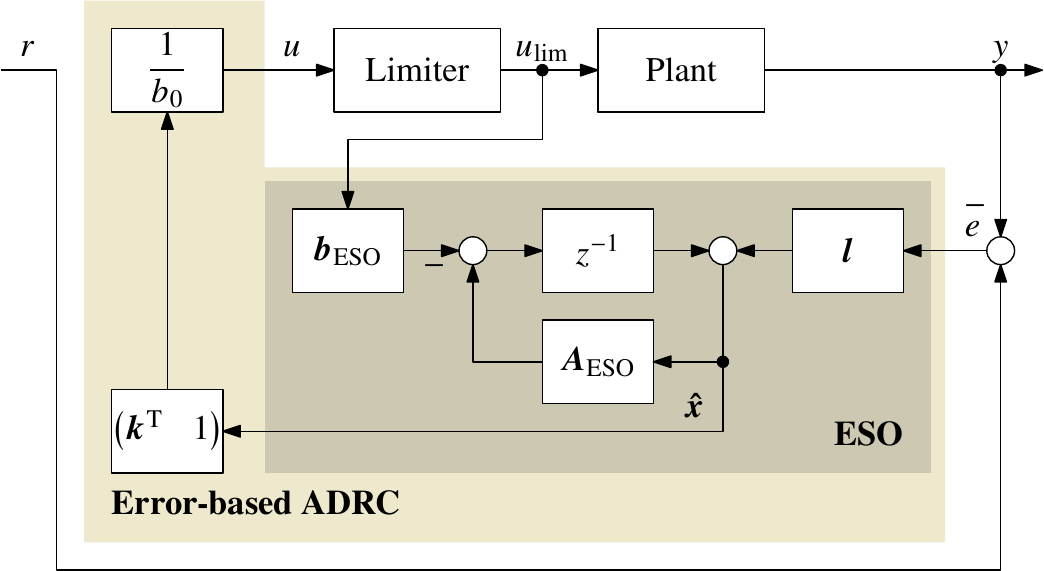}
    \caption{Control loop with discrete-time error-based ADRC, state space implementation}
    \label{fig:ADRC_DTe_SS_Limiter}
\end{figure}

This enables us to take a very convenient shortcut.
We can simply reuse the discrete-time observer equation \refEq{eqn:ADRC_DT_SS_Observer}
from \refSec{sec:ADRC_DT_ControlLaw}, exchange $y(k)$ for $e(k)$, and change the sign for $u_\mathrm{lim}(k)$.
The properly discretized observer for error-based ADRC now reads:
\begin{equation}
    \Vector{\hat{x}}(k) = \Matrix{A}_\mathrm{ESO} \cdot \Vector{\hat{x}}(k-1) - \Vector{b}_\mathrm{ESO} \cdot u_\mathrm{lim}(k-1) + \Vector{l} \cdot e(k)
    ,
    \label{eqn:ADRC_DTe_SS_Observer}
\end{equation}
with $\Matrix{A}_\mathrm{ESO}$ and $\Vector{b}_\mathrm{ESO}$ as well as controller/observer gains being the same as in the output-based case.
The block diagram shown in \refFig{fig:ADRC_DTe_SS_Limiter} is, consequently, also very similar to \refFig{fig:ADRC_DT_SS_Limiter}.

Please remember that even though this derivation could be presented in an extremely condensed form, it is the first discrete-time implementation of error-based ADRC in state space.
Owing to the equivalence to output-based ADRC at the core observer dynamics, all findings of \cite{Miklosovic:2006} regarding the best discretization strategy for linear ADRC automatically apply here as well (zero-order hold discretization in conjunction with the current observer approach).

This also means that there is no need for a ``custom'' discretization of error-based ADRC at transfer function level, as done in \cite{Madonski:2019}---we can instead follow the same route of exact transfer function representations as presented in \refSec{sec:DT_TF} of this article, and will do so right next.

\subsection{Discrete-Time Transfer Function Implementation}
\label{sec:ErrorBased_DT_TF}

As the counterpart of the output-based version in \refSec{sec:DT_TF_1}, a transfer function implementation of discrete-time error-based ADRC can now easily be derived by following the same steps.
We once again start with the $z$-domain representation of control law \refEq{eqn:ADRC_DTe_SS_Controller} and observer \refEq{eqn:ADRC_DTe_SS_Observer}, and use $u(z) = u_\mathrm{lim}(z)$ as an observer input, as we will lose the dedicated feedback path of the limited controller output:
\begin{gather}
    u(z) = \frac{1}{b_0} \cdot \begin{pmatrix} \Vector{k}^\Transpose & 1 \end{pmatrix} \cdot \Vector{\hat{x}}(z)
    \label{eqn:ADRC_DTe_TF_ControlLaw_z}
    ,\\
    \Vector{\hat{x}}(z) = z^{-1} \cdot \Matrix{A}_\mathrm{ESO} \cdot \Vector{\hat{x}}(z) - z^{-1} \cdot \Vector{b}_\mathrm{ESO} \cdot u(z) + \Vector{l} \cdot e(z)
    .
    \label{eqn:ADRC_DTe_TF_Observer_z}
\end{gather}

\begin{figure}
    \includegraphics[scale=0.8]{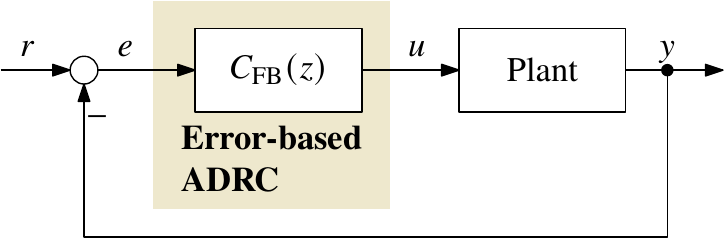}
    \caption{Control loop with discrete-time error-based ADRC, transfer function implementation}
    \label{fig:ADRC_DTe_TF}
\end{figure}

The closed-loop observer equation is obtained by putting \refEq{eqn:ADRC_DTe_TF_ControlLaw_z} into \refEq{eqn:ADRC_DTe_TF_Observer_z}:
\begin{equation*}
    \Vector{\hat{x}}(z)
    = \Matrix{\Phi}_\mathrm{ESO} \cdot \Vector{l} \cdot e(z)
    ,
\end{equation*}
where $\Matrix{\Phi}_\mathrm{ESO}$ is identical to \refEq{eqn:ADRC_DT_TF_Observer_Phi} in the output-based case.
Inserting this equation back into \refEq{eqn:ADRC_DTe_TF_ControlLaw_z} yields the control law for the transfer function implementation of error-based ADRC, which is also shown in \refFig{fig:ADRC_DTe_TF}:
\begin{equation}
    u(z) = \underbrace{
        \frac{1}{b_0} \cdot \begin{pmatrix} \Vector{k}^\Transpose & 1 \end{pmatrix}
        \cdot \Matrix{\Phi}_\mathrm{ESO} \cdot \Vector{l}
    }_{C_\mathrm{FB}(z)}
    \cdot e(z)
    .
    \label{eqn:ADRC_DTe_TF_ControlLaw}
\end{equation}

Equivalent to the continuous-time derivation in \cite{Madonski:2023a}, only one transfer function is required here.
And, above all, exactly the same transfer function $C_\mathrm{FB}(z)$ given in \refEq{eqn:ADRC_DT_TF_CFB} is being used for output- and error-based ADRC, with the same $\alpha$ and $\beta$ coefficients from \refTable{table:DT_TF_Coefficients_1} and \refTable{table:DT_TF_Coefficients_2}.

Comparing control laws and block diagrams of discrete-time output- and error-based ADRC in \refFig{fig:ADRC_DT_TF} and \refFig{fig:ADRC_DTe_TF}, it is clear that output-based ADRC can be seen as an extension of error-based ADRC, with an additional reference signal prefilter $C_\mathrm{PF}(z)$---as in the continuous-time case \cite{Madonski:2023a}.

In output-based ADRC, the dynamics of reference tracking and disturbance rejection cannot be independently chosen, the prefilter is tuned for a critically damped closed-loop response.
Being a more ``bare-bones'' version, error-based ADRC therefore allows to be more flexibly combined with an independently tuned prefilter for an actual two-degrees-of-freedom design.

\subsection{Discrete-Time Dual-Feedback Transfer Function Implementation}
\label{sec:ErrorBased_DT_FBTF}

\begin{figure}
    \includegraphics[width=\linewidth]{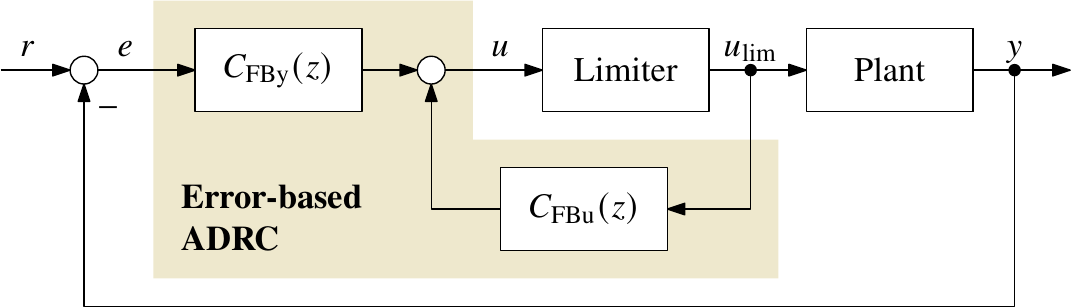}
    \caption{Control loop with discrete-time error-based ADRC, dual-feedback transfer function implementation}
    \label{fig:ADRC_DTe_FBTF}
\end{figure}

\begin{figure*}
    \subfloat[Output-based, first-order]{%
        \includegraphics[scale=0.65]{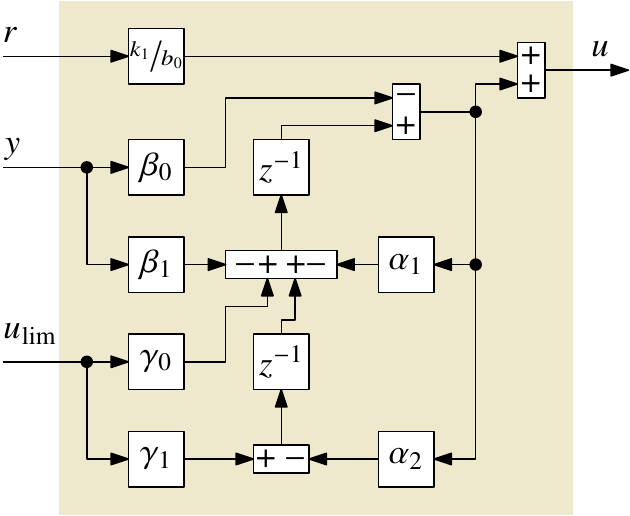}%
        \label{fig:ADRC_DT_FBTF_1}
    }%
    \hfill%
    \subfloat[Output-based, second-order]{%
        \includegraphics[scale=0.65]{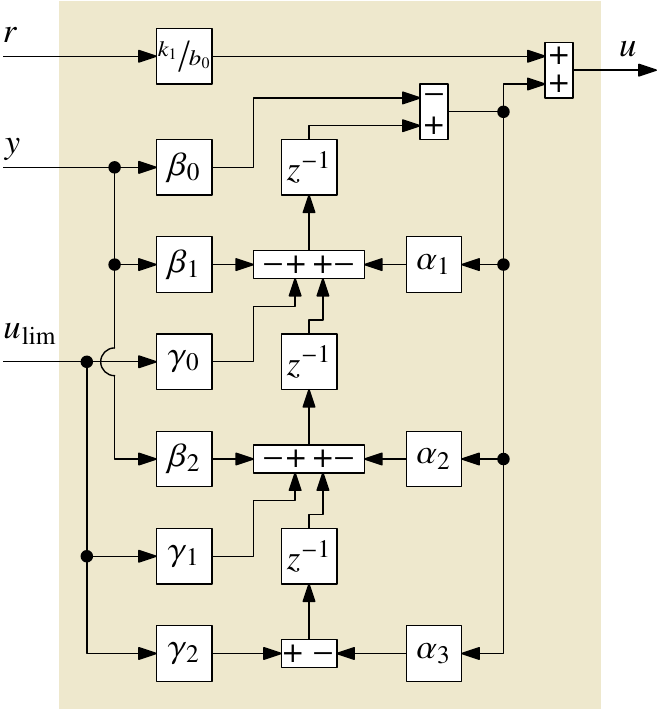}%
        \label{fig:ADRC_DT_FBTF_2}
    }%
    \hfill%
    \subfloat[Error-based, first-order]{%
        \includegraphics[scale=0.65]{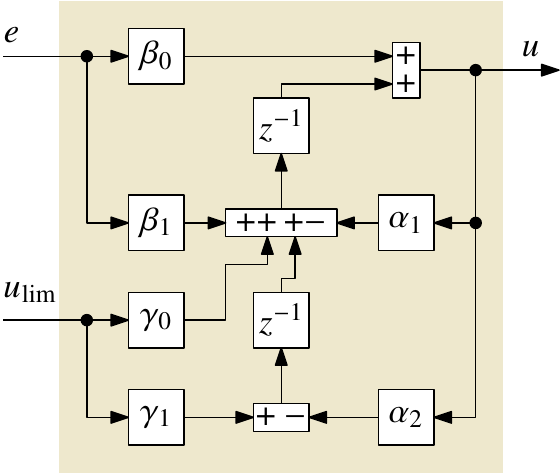}%
        \label{fig:ADRC_DTe_FBTF_1}
    }%
    \hfill%
    \subfloat[Error-based, second-order]{%
        \includegraphics[scale=0.65]{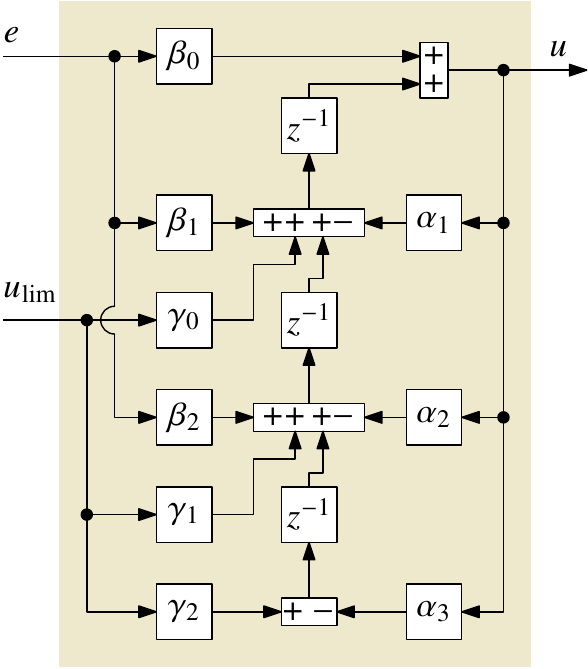}%
        \label{fig:ADRC_DTe_FBTF_2}
    }%
    \caption{Dual-feedback transfer function implementations of discrete-time ADRC variants with minimum footprint}
\end{figure*}

With the discrete-time state-space implementation of error-based ADRC in \refSec{sec:ErrorBased_DT} at hand, a dual-feedback transfer function variant can easily be derived, as well.
The $z$-transform of the observer \refEq{eqn:ADRC_DTe_SS_Observer},
\begin{equation*}
    \Vector{\hat{x}}(z) = z^{-1} \cdot \Matrix{A}_\mathrm{ESO} \cdot \Vector{\hat{x}}(z) - z^{-1} \cdot \Vector{b}_\mathrm{ESO} \cdot u_\mathrm{lim}(z) + \Vector{l} \cdot e(z)
    ,
\end{equation*}
is being solved for $\Vector{\hat{x}}(z)$, yielding:
\begin{equation*}
    \begin{aligned}
        \Vector{\hat{x}}(z) = & \left( \Matrix{I} - z^{-1} \Matrix{A}_\mathrm{ESO} \right)^{-1}
        \\
        & \cdot \left(
        - z^{-1} \cdot \Vector{b}_\mathrm{ESO} \cdot u_\mathrm{lim}(z) + \Vector{l} \cdot e(z)
        \right)
        ,
    \end{aligned}
\end{equation*}
which can be inserted into the control law \refEq{eqn:ADRC_DTe_TF_ControlLaw_z}:
\begin{equation}
    \begin{aligned}
        u(z) = & \frac{1}{b_0} \cdot \begin{pmatrix} \Vector{k}^\Transpose & 1 \end{pmatrix}
        \cdot \left( \Matrix{I} - z^{-1} \Matrix{A}_\mathrm{ESO} \right)^{-1}
        \\
        & \cdot \left(
        - z^{-1} \cdot \Vector{b}_\mathrm{ESO} \cdot u_\mathrm{lim}(z) + \Vector{l} \cdot e(z)
        \right)
        .
    \end{aligned}
    \label{eqn:ADRC_DTe_FBTF_Controller_Detail}
\end{equation}

By inspection and comparison with the output-based counterpart \refEq{eqn:ADRC_DT_FBTF_Controller}. we can make the important observation that \refEq{eqn:ADRC_DTe_FBTF_Controller_Detail} can be represented using the two already introduced transfer functions $C_\mathrm{FBy}(z)$ and $C_\mathrm{FBu}(z)$:
\begin{equation}
    u(z) = C_\mathrm{FBy}(z) \cdot e(z) + C_\mathrm{FBu}(z) \cdot u_\mathrm{lim}(z)
    ,
    \label{eqn:ADRC_DTe_FBTF_Controller}
\end{equation}
where $C_\mathrm{FBy}(z)$ and $C_\mathrm{FBu}(z)$ are identical to their definition in \refEq{eqn:ADRC_DT_FBTF_FBy} and \refEq{eqn:ADRC_DT_FBTF_FBu}.
The required $\alpha$, $\beta$, $\gamma$ coefficients can, consequently, be reused in unmodified form as given in \refTable{table:DT_FBTF_Coefficients_1} and \refTable{table:DT_FBTF_Coefficients_2} for the first- and second-order case.

Please note that we kept the subscript ``y'' as part of $C_\mathrm{FBy}(z)$ in the control law  \refEq{eqn:ADRC_DTe_FBTF_Controller} and its visualization in \refFig{fig:ADRC_DTe_FBTF} to emphasize that exactly the same transfer functions from the output-based variant are being reused here, even though $C_\mathrm{FBy}(z)$ is now being fed with the control error $e$---which, in turn, of course also contains $y$.

As a final step, a very efficient implementation of this control law can be found as well using the ideas from \cite{Herbst:2021b}.
With the common denominator of $C_\mathrm{FBu}(z)$ and $C_\mathrm{FBy}(z)$, the control law \refEq{eqn:ADRC_DTe_FBTF_Controller} can be written as:
\begin{equation}
    u(z) = \frac{
        z^{-1} \cdot \left( \displaystyle\sum_{i = 0}^{n} \beta_i z^{-i } \right) \cdot u_\mathrm{lim}(z)
        + \left( \displaystyle\sum_{i = 0}^{n} \gamma_i z^{-i} \right) \cdot e(z)
    }{
        1 + \displaystyle\sum_{i = 1}^{n+1} \alpha_i z^{-i}
    }
    .
    \label{eqn:ADRC_DTe_FBTF_ControlLaw_Superposition}
\end{equation}

A direct form II implementation of \refEq{eqn:ADRC_DTe_FBTF_ControlLaw_Superposition} reduces both storage variables and algebraic operations required during runtime to a minimum.
The resulting implementations are shown in \refFig{fig:ADRC_DTe_FBTF_1} and \refFig{fig:ADRC_DTe_FBTF_2} for the first- and second-order case, allowing a direct comparison with their output-based counterparts.

\section{Example}
\label{sec:Example}

In this section, a power electronics application (DC-DC converter) to be controlled by first-order output- and error-based ADRC is being studied.%
\footnote{%
    For output-based ADRC, this is a replication of the example in \cite{Herbst:2021b}.
}
As we developed error-based ADRC implementations in this paper that exhibit exactly the same dynamic disturbance rejection behavior as their output-based counterparts, the equivalence between output- and error-based ADRC in this regard should therefore be visible.
Furthermore, the anti-windup abilities shall be demonstrated that error-based ADRC gained in the implementations introduced in this paper by keeping the separate feedback path for the limited controller output in both the state-space and the dual-feedback variant.
For comparability down to the level of individual sampling intervals, we will perform this comparison in a simulation environment.

\begin{figure}%
    \centering%
    \includegraphics[width=\linewidth]{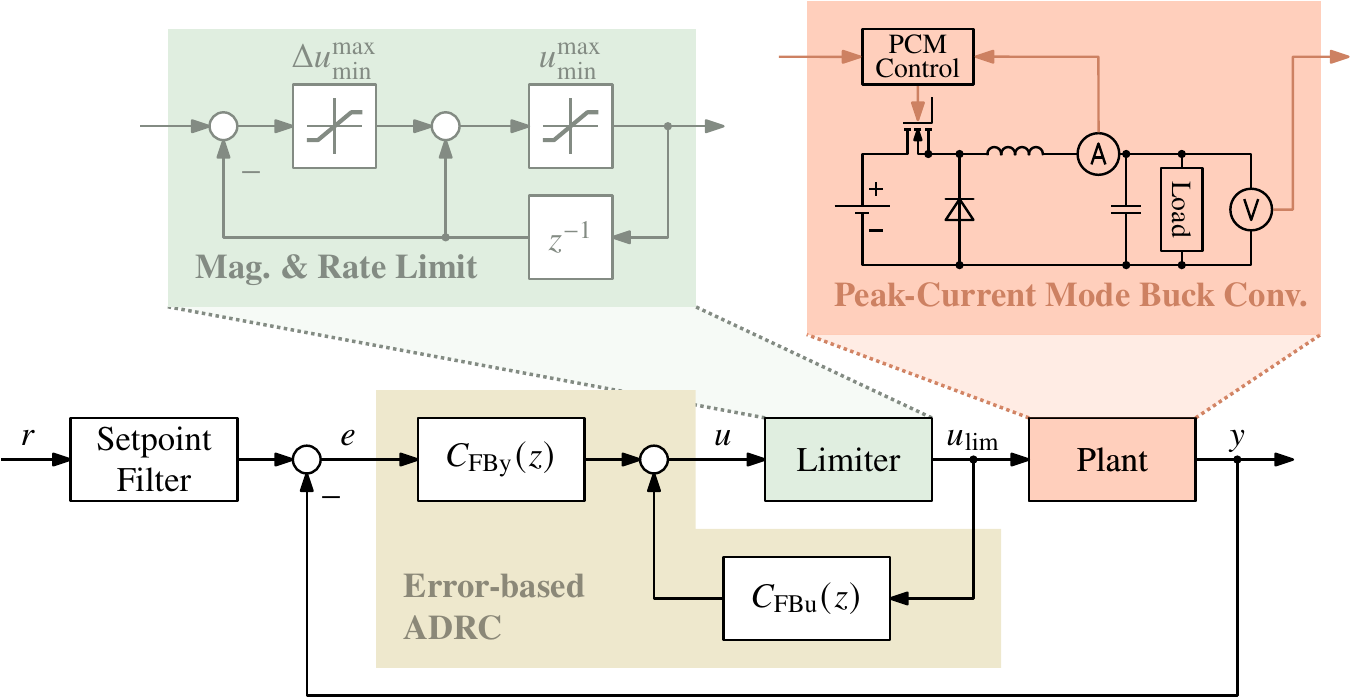}
    \caption{%
        Schematic of a voltage control loop around a peak-current mode controlled step-down DC-DC converter.
        Discrete-time error-based ADRC in the dual-feedback transfer function implementation is being used as voltage controller, enhanced by a first-order low-pass filter to adjust the setpoint tracking dynamics.
        To limit the absolute value as well as the slope of the converter current during transients, the voltage controller is equipped with a rate and magnitude limitation block.
    }
    \label{fig:Example_Schematic}
\end{figure}

\begin{figure*}%
    \centering%
    \includegraphics[scale=1.0,page=1]{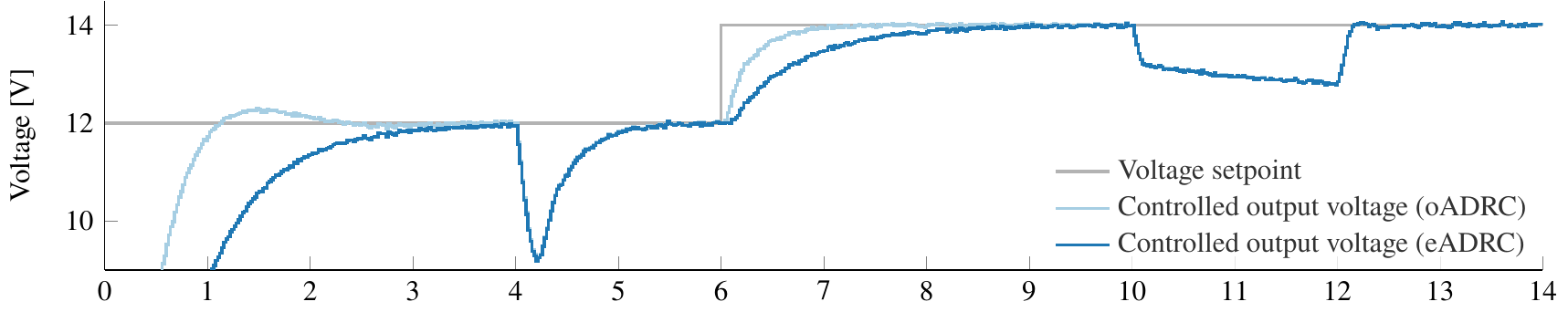}\\%
    \includegraphics[scale=1.0,page=2]{Demo_ADRC_BuckPCM}\\%
    \includegraphics[scale=1.0,page=3]{Demo_ADRC_BuckPCM}\\%
    \includegraphics[scale=1.0,page=4]{Demo_ADRC_BuckPCM}%
    \caption{%
        Simulation results for the application example from \refFig{fig:Example_Schematic}.
        In each diagram, the result of output-based ADRC (oADRC) is plotted with brighter, error-based ADRC (eADRC) with darker color.
        On can very well spot the identical behavior of oADRC and eADRC during load transients and when returning from saturation---no matter if limited by magnitude or by rate (active limitation is shown in the bottom diagram).
        As to be expected, only the reference tracking behavior differs.
        The eADRC variant was equipped with an additional setpoint filter for a slower, more smooth response in a true two-degrees-of-freedom design.
        Note that the voltage drop between $10\,\mathrm{ms} < t < 12\,\mathrm{ms}$ does not result from the controllers inability to reject disturbances, but from the load current being (deliberately chosen) so high that the input current the (saturated) controller is allowed to supply does not suffice to prevent the capacitor from discharging.
    }
    \label{fig:Example_Results}
\end{figure*}

An overview of the DC-DC converter and its control structure with error-based ADRC is shown in \refFig{fig:Example_Schematic}.
The converter parameters are:
$V_\mathrm{Input} = 24\,\mathrm{V}$,
$L = 33\,\mathrm{\upmu H}$,
$C = 100\,\mathrm{\upmu F}$,
base load $R = 100\,\mathrm{\Omega}$,
$f_\mathrm{Switch} = 50\,\mathrm{kHz} = \frac{1}{T}$,
peak-current mode slope compensation of $2\,\mathrm{A}$ per cycle.
To limit not only the absolute values, but also the slope of the converter current during transients, the voltage controller is equipped with a magnitude and rate limitation block configured for peak current commands from the interval $[0\,\mathrm{A}, 6\,\mathrm{A}]$ with a rate limit of $\pm 20\,\mathrm{A}/\mathrm{ms}$.
To dynamically decouple reference tracking from disturbance rejection, a first-order low-pass setpoint filter is added, with a time constant of $750\,\mathrm{\upmu s}$ for a $98\,\%$ settling time of $3\,\mathrm{ms}$.
ADRC tuning is performed with a bandwidth $\omega_\mathrm{CL} = 4000\,\mathrm{rad/s}$ for an $98\,\%$ settling time of $1\,\mathrm{ms}$, and a relative observer bandwidth $k_\mathrm{ESO} = 5$.
As shown in \cite{Herbst:2016a}, plant modeling for ADRC is extremely easy in this application, depending only on the output capacitor with $b_0 = \frac{1}{C} = 10000\,\mathrm{V/As}$.
The sampling frequency of the discrete-time implementation corresponds to the switching frequency.
To be closer to practice, measurement noise with $\sigma = 20\,\mathrm{mV}$ is added to the output voltage, and a full sampling interval assumed as the delay between measurement and updated controller output.

The scenario consists of startup at (almost) no load, a jump to full load at $t = 4\,\mathrm{ms}$, a setpoint change at $t = 6\,\mathrm{ms}$, and a temporary overload condition between $t = 10\,\mathrm{ms}$ and $t = 12\,\mathrm{ms}$---the latter to deliberately drive the controller into saturation.
The simulation is performed twice: once with error-based ADRC (including the setpoint filter), and once with output-based ADRC (without setpoint filter).
Results are presented in \refFig{fig:Example_Results}.

The simulation results confirm that the discrete-time error-based implementations introduced in this article deliver identical disturbance rejection behavior as their output-based counterparts, down to the scope of individual samples.
As an additional plus, they can now benefit from the same windup protection options.

\section{Conclusion}

\begin{table*}
    \caption{%
        Overview of discrete-time implementations for output-based and error-based ADRC presented in this article
    }
    \label{table:VariantOverview}
    \begin{tabular*}{\textwidth}{@{\extracolsep\fill}lllp{\columnwidth}@{\extracolsep\fill}}
        \toprule
        \textbf{Implementation}  &  \textbf{Output-based}  &  \textbf{Error-based}  &  \textbf{Characteristics}  \\
        \midrule
        State space
            &   \refSec{sec:ADRC_DT_ControlLaw}
            &   \refSec{sec:ErrorBased_DT}
            &   Discrete-time equivalent of the original structure of linear continuous-time ADRC.
                Flexible control signal limitation and windup protection.
                Runtime: Minimum number of storage variables,
                maximum number of coefficients and operations.
        \\
        Transfer function
            &   \refSec{sec:DT_TF_1}
            &   \refSec{sec:ErrorBased_DT_TF}
            &   Strong similarity to classical digital control using feedback controller and prefilter.
                Control signal limitation only via clamped integrator.
                Runtime: Maximum number of storage variables,
                reduced number of coefficients and operations.
        \\
        Dual-feedback TF
            &   \refSec{sec:DT_TF_2}
            &   \refSec{sec:ErrorBased_DT_FBTF}
            &   Full replication of the state-space behavior using transfer functions, but with uncommon structure.
                Flexible control signal limitation and windup protection.
                Runtime: Minimum number of storage variables, coefficients, and operations.
        \\
        \bottomrule
    \end{tabular*}
\end{table*}

Practical implementations of ADRC are usually carried out in software, i.\,e.\ in a discrete time domain.
This article provided both a summary and an extension of discrete-time implementations of linear ADRC.
As a first contribution, controller tuning was performed using bandwidth parameterization with discrete-time pole placement, departing from the prevalent quasi-continuous design procedure.
A bridge to the existing tuning rules could be built, while also demonstrating the necessity of accurate results for low sampling frequencies.
For two existing discrete-time transfer function implementations of ADRC, coefficients were rederived based on accurate discrete-time pole placement.

A second bridge was built from ``classical'' (output-based) to error-based ADRC in the discrete time domain.
It was shown that existing implementations need only slight restructuring, and transfer functions, coefficients, tuning rules can all be reused for error-based ADRC.
In one sweep, three new error-based ADRC implementations could be derived in this article, which are all direct counterparts of existing output-based variants.
\refTableBegin{table:VariantOverview} summarizes all variants presented here along with some main characteristics, to assist with making a choice.

Error- and output-based implementations now exhibit exactly the same inner-loop behavior (disturbance rejection), rendering custom discrete-time implementations of error-based ADRC---as done e.\,g.\ in \cite{Madonski:2019}---unnecessary.
As an important feature for real-world applications, error- and output-based ADRC can now be equipped with the same control signal limitation and windup protection capabilities.

Be it the necessity of a true two-degrees-of-freedom design, windup protection, a low footprint during runtime, or structural similarity to solutions of classical digital control---the discrete-time ADRC variants introduced and discussed in this article can now cater a larger variety of practical needs than ever.
What a great time to put ADRC into practice!



\begin{thebibliography}{10}

\bibitem{Han:2009}
J.~Han, ``From {PID} to active disturbance rejection control,'' \emph{IEEE
  Transactions on Industrial Electronics}, vol.~56, no.~3, pp. 900--906, 2009.

\bibitem{Zheng:2010}
Q.~Zheng and Z.~Gao, ``On practical applications of active disturbance
  rejection control,'' in \emph{Proceedings of the 29th Chinese Control
  Conference}, 2010, pp. 6095--6100.

\bibitem{Zheng:2018}
------, ``Active disturbance rejection control: Some recent experimental and
  industrial case studies,'' \emph{Control Theory and Technology}, vol.~16,
  no.~4, pp. 301--313, 2018.

\bibitem{Talole:2018}
S.~E. Talole, ``Active disturbance rejection control: {Applications} in
  aerospace,'' \emph{Control Theory and Technology}, vol.~16, pp. 314--323,
  2018.

\bibitem{Farah:2021}
R.~Fareh, S.~Khadraoui, M.~Y. Abdallah, M.~Baziyad, and M.~Bettayeb, ``Active
  disturbance rejection control for robotic systems: {A} review,''
  \emph{Mechatronics}, vol.~80, p. 102671, 2021.

\bibitem{Gao:2006}
Z.~Gao, ``Active disturbance rejection control: {A} paradigm shift in feedback
  control system design,'' in \emph{Proceedings of the 2006 American Control
  Conference}, 2006, pp. 2399--2405.

\bibitem{Gao:2003}
------, ``Scaling and bandwidth-parameterization based controller tuning,'' in
  \emph{Proceedings of the 2003 American Control Conference}, 2003, pp.
  4989--4996.

\bibitem{Miklosovic:2006}
R.~Miklosovic, A.~Radke, and Z.~Gao, ``Discrete implementation and
  generalization of the extended state observer,'' in \emph{Proceedings of the
  2006 American Control Conference}, 2006, pp. 2209--2214.

\bibitem{Madonski:2015}
R.~Madonski, Z.~Gao, and K.~\L{}akomy, ``Towards a turnkey solution of
  industrial control under the active disturbance rejection paradigm,'' in
  \emph{2015 54th Annual Conference of the Society of Instrument and Control
  Engineers of Japan (SICE)}, 2015, pp. 616--621.

\bibitem{Herbst:2016a}
G.~Herbst, ``Practical active disturbance rejection control: Bumpless transfer,
  rate limitation, and incremental algorithm,'' \emph{IEEE Transactions on
  Industrial Electronics}, vol.~63, no.~3, pp. 1754--1762, 2016.

\bibitem{Madonski:2015b}
R.~Madonski and P.~Herman, ``Survey on methods of increasing the efficiency of
  extended state disturbance observers,'' \emph{ISA Transactions}, vol.~56, pp.
  18--27, 2015.

\bibitem{Fu:2016}
C.~Fu and W.~Tan, ``Tuning of linear {ADRC} with known plant information,''
  \emph{ISA Transactions}, vol.~65, pp. 384--393, 2016.

\bibitem{Zhou:2019}
R.~Zhou and W.~Tan, ``Analysis and tuning of general linear active disturbance
  rejection controllers,'' \emph{IEEE Transactions on Industrial Electronics},
  vol.~66, no.~7, pp. 5497--5507, 2019.

\bibitem{Tian:2007}
G.~Tian and Z.~Gao, ``Frequency response analysis of active disturbance
  rejection based control system,'' in \emph{2007 IEEE International Conference
  on Control Applications}, 2007, pp. 1595--1599.

\bibitem{Huang:2013:CCC}
C.~Huang and Z.~Gao, ``On transfer function representation and frequency
  response of linear active disturbance rejection control,'' in
  \emph{Proceedings of the 32th Chinese Control Conference}, 2013, pp. 72--77.

\bibitem{Zheng:2016}
Q.~Zheng and Z.~Gao, ``Active disturbance rejection control: Between the
  formulation in time and the understanding in frequency,'' \emph{Control
  Theory and Technology}, vol.~14, no.~3, pp. 250--259, 2016.

\bibitem{Herbst:2021a}
G.~Herbst, ``Transfer function analysis and implementation of active
  disturbance rejection control,'' \emph{Control Theory and Technology},
  vol.~19, pp. 19--34, 2021.

\bibitem{Herbst:2021b}
------, ``A minimum-footprint implementation of discrete-time {ADRC},'' in
  \emph{2021 European Control Conference (ECC)}, 2021, pp. 107--112.

\bibitem{Michalek:2016}
M.~M. Micha\l{}ek, ``Robust trajectory following without availability of the
  reference time-derivatives in the control scheme with active disturbance
  rejection,'' in \emph{Proceedings of the 2016 American Control Conference},
  2016, pp. 1536--1541.

\bibitem{Mandali:2020}
A.~Mandali, L.~Dong, and A.~Morinec, ``Robust controller design for automatic
  voltage regulation,'' in \emph{Proceedings of the 2020 American Control
  Conference}, 2020, pp. 2617--2622.

\bibitem{Lechekhab:2021}
T.~E. Lechekhab, S.~M. Manojlovi\'{c}, M.~R. Stankovi\'{c}, R.~Madonski, and
  S.~M. Simi\'{c}, ``Robust error-based active disturbance rejection control of
  a quadrotor,'' \emph{Aircraft Engineering and Aerospace Technology}, vol.~93,
  no.~1, pp. 89--104, 2021.

\bibitem{Madonski:2021}
R.~Madonski, K.~\L{}akomy, and J.~Yang, ``Simplifying {ADRC} design with
  error-based framework: Case study of a {DC-DC} buck power converter,''
  \emph{Control Theory and Technology}, vol.~19, pp. 94--112, 2021.

\bibitem{Huang:2022}
T.~Huang, G.~Hu, Y.~Yan, D.~Zeng, and Z.~Meng, ``Combined feedforward and
  error-based active disturbance rejection control for diesel particulate
  filter thermal regeneration,'' \emph{ISA Transactions}, 2022.

\bibitem{Madonski:2023a}
R.~Madonski, G.~Herbst, and M.~R. Stankovi\'{c}, ``{ADRC} in output and error
  form: {Connection}, equivalence, performance,'' \emph{Control Theory and
  Technology}, vol.~21, pp. 56--71, 2023.

\bibitem{Madonski:2019}
R.~Madonski, S.~Shao, H.~Zhang, Z.~Gao, J.~Yang, and S.~Li, ``General
  error-based active disturbance rejection control for swift industrial
  implementations,'' \emph{Control Engineering Practice}, vol.~84, pp.
  218--229, 2019.

\bibitem{AstromMurray:2021}
K.~J. \r{A}str{\"o}m and R.~M. Murray, \emph{Feedback Systems: {An}
  Introduction for Scientists and Engineers}, 2nd~ed.\hskip 1em plus 0.5em
  minus 0.4em\relax Princeton, NJ, USA: Princeton University Press, 2021.

\bibitem{Herbst:2013}
G.~Herbst, ``A simulative study on active disturbance rejection control
  ({ADRC}) as a control tool for practitioners,'' \emph{Electronics}, vol.~2,
  no.~3, pp. 246--279, 2013.

\bibitem{Franklin:1997}
G.~F. Franklin, M.~L. Workman, and D.~Powell, \emph{Digital Control of Dynamic
  Systems}, 3rd~ed.\hskip 1em plus 0.5em minus 0.4em\relax Boston, MA, USA:
  Addison-Wesley Longman Publishing, 1997.

\bibitem{Miklosovic:2007}
R.~Miklosovic and A.~Radke, ``High performance tracking control for the
  practitioner,'' in \emph{2007 American Control Conference}, 2007, pp.
  3009--3014.

\bibitem{Zhang:2013}
Y.~Zhang, Y.~Zhang, J.~Wang, and R.~Ma, ``An active disturbance rejection
  control of induction motor using {DSP}+{FPGA},'' in \emph{2013 25th Chinese
  Control and Decision Conference (CCDC)}, 2013, pp. 4047--4052.

\bibitem{Stankovic:2016}
M.~R. Stankovi\'{c}, S.~M. Manojlovi\'{c}, S.~M. Simi\'{c}, S.~T. Mitrovi\'{c},
  and M.~B. Naumovi\'{c}, ``{FPGA} system-level based design of multi-axis
  {ADRC} controller,'' \emph{Mechatronics}, vol.~40, pp. 146--155, 2016.

\bibitem{Ahi:2018}
B.~Ahi and A.~Nobakhti, ``Hardware implementation of an {ADRC} controller on a
  gimbal mechanism,'' \emph{IEEE Transactions on Control Systems Technology},
  vol.~26, no.~6, pp. 2268--2275, 2018.

\bibitem{Desai:2018}
R.~Desai, B.~M. Patre, and S.~N. Pawar, ``Active disturbance rejection control
  with adaptive rate limitation for process control application,'' in
  \emph{2018 Indian Control Conference (ICC)}, 2018, pp. 131--136.

\bibitem{RamirezNeria:2020}
M.~Ram{\'i}rez-Neria, A.~Luviano-Ju{\'a}rez, N.~Lozada-Castillo,
  G.~Ochoa-Ortega, and R.~Madonski, ``Discrete-time active disturbance
  rejection control: {A} delta operator approach,'' in \emph{Advanced,
  Contemporary Control}, A.~Bartoszewicz, J.~Kabzi{\'{n}}ski, and J.~Kacprzyk,
  Eds.\hskip 1em plus 0.5em minus 0.4em\relax Cham, Switzerland: Springer
  International Publishing, 2020, pp. 1383--1395.

\bibitem{Peng:1996}
Y.~Peng, D.~Vran\v{c}i\v{c}, and R.~Hanus, ``Anti-windup, bumpless, and
  conditioned transfer techniques for {PID} controllers,'' \emph{IEEE Control
  Systems Magazine}, vol.~16, no.~4, pp. 48--57, 1996.

\bibitem{Astrom:2006}
K.~J. \r{A}str{\"o}m and T.~H\"{a}gglund, \emph{Advanced {PID} Control}.\hskip
  1em plus 0.5em minus 0.4em\relax Durham, NC, USA: ISA---The Instrumentation,
  Systems, and Automation Society, 2006.

\end{thebibliography}
\end{document}